\documentclass[varenna]{cimento}

\usepackage{graphicx}
\usepackage{bbm}
\usepackage{bm}
\usepackage[dvipsnames]{xcolor}
\usepackage{amssymb,amsmath}
\usepackage{bbold}
\usepackage[normalem]{ulem}
\usepackage{booktabs}
\usepackage{longtable}
\usepackage{braket}
\usepackage[T1]{fontenc}
\usepackage[utf8]{inputenc}
\usepackage[english]{babel}
\usepackage{xspace}
\usepackage{epstopdf}
\usepackage[final]{microtype}
\usepackage{ragged2e}
\usepackage{mcite}
\usepackage{cite}

\renewcommand{\vec}[1]{\boldsymbol{#1}}

\def\up{\uparrow}

\def\kk{\vec{k}}

\def\qq{\vec{q}}

\def\tt{\mathbf{t}}

\def\MR{moir\'e}

\newcommand{\qql}{\textquotedblleft}
\newcommand{\qqr}{\textquotedblright\xspace}
\newcommand{\vc}[1]{\vec{#1}} 

\graphicspath{{./Figures/}}

\newcommand{\crea}[1]{#1^{\dag}}
\newcommand{\ani}[1]{#1^{\vphantom{\dag}}}
\newcommand{\ave}[1]{\left\langle#1\right\rangle}
\newcommand{\dn}{\downarrow}

\title{Quantum geometry in superfluidity\\ and superconductivity}

\author{Sebastiano Peotta, Kukka-Emilia Huhtinen, \atque P\"aivi T\"orm\"a\thanks{E-mail: paivi.torma@aalto.fi}}
\institute{Department of Applied Physics, Aalto University School of Science,\\ FI-00076 Aalto, Finland}

\begin{document}

\maketitle

\vspace{-1.8cm}

\begin{abstract}
We review the theoretical description of the role of quantum geometry in superfluidity and superconductivity of multiband systems, with focus on flat bands where quantum geometry is wholly responsible for supercurrents. This review differs from previous ones in that it is based on the most recent understanding of the theory: the dependence of the self-consistent order parameter on the supercurrent is properly taken into account, and the superfluid weight in a flat band becomes proportional to the minimal quantum metric. We provide a recap of basic quantum geometric quantities and the concept of superfluid density. The geometric contribution of superconductivity is introduced via considering the two-body problem. The superfluid weight of a multiband system is derived within mean-field theory, leading to a topological bound of flat band superconductivity. The physical interpretation of the flat band supercurrent in terms of Wannier function overlaps is discussed.
\end{abstract}

\section{Introduction}
\label{sec:introduction}

The basic theory of superfluidity and superconductivity is based on Bose-Einstein condensation (BEC) of bosonic particles. In the case of superconductivity, these bosons are Cooper pairs of fermions of two opposite spins. For weak inter-particle interactions, the condensate of Cooper pairs is described by the Bardeen-Cooper-Schrieffer (BCS) mean-field theory of superconductivity~\cite{Schrieffer1964}, while very tightly bound pairs condense similarly to non-composite bosons --- these two regimes are connected by the BEC-BCS crossover~\cite{zwerger2012}. The dispersion and dimensionality of the system determines whether BEC is possible in the thermodynamic limit, but otherwise the condensation of particles in free space, trapped systems, and single band lattices is quite similar. Qualitatively new phenomena arise, however, when the system has more degrees of freedom: multicomponent condensates and boson-fermion mixtures are examples of this.  Another type of complexity arises when the particles are confined in lattices where the unit cell has multiple orbitals (lattice sites). In this case, multiband extensions of the theory of superfluidity and superconductivity must be applied. Recently, it has been shown that superfluidity and superconductivity in a multiband system has an intriguing connection to quantum geometry, which is particularly relevant in flat (dispersionless) bands.

Superconductivity and fermionic superfluidity require Cooper pair formation of electrons of opposite spins, or of the two components of an ultracold two-component Fermi gas. According to the BCS mean-field theory, $T_c \propto e^{-{1/(\rho(E_F) |U|)}}$, where $\rho(E_F)$ is the density of states at the Fermi level, and $U$ the contact (on-site) interaction between the two spins (two components). While this tells about Cooper pairing, to show that the system is actually superconducting, or a neutral superfluid, one must prove that there can be dissipationless supercurrent. For that, the superfluid weight $D_s$ needs to be determined. In a single band system, it is typically proportional to the inverse effective mass of the band (second derivative of the dispersion with respect to the lattice momentum). Essentially, the superfluid weight tells that superfluids made of heavy particles have weaker supercurrents, which is intuitive. Remarkably, however, it has been recently shown that in the case of multiband superconductivity, $D_s$ contains an additional contribution beyond the inverse band effective mass: a so-called geometric contribution~\cite{Peotta2015,Julku2016,Liang2017,Torma2018,Huhtinen2022}. Geometric here refers to the distances between quantum states.  

The geometric contribution of superconductivity may be highly relevant in the search of new superconductors with higher critical temperature. In bands where the energy dispersion as function of momentum, $\epsilon(\vc{k})$, is constant -- so-called flat bands (line-graph lattices, split graph, Lieb, etc.~\cite{mielke1991a, lieb1989,Leykam2018}), the critical temperature $T_{c}$ at which Cooper pairs form is predicted~\cite{Kopnin2011,Heikkila2011,Khodel1994} to be linearly proportional to the interaction $U$. Comparing this to the above usual (single-band) BCS formula, one can see that the exponential suppression of $T_c$ for small interactions no longer exists. This gives great promise for high temperature superconductivity. However, in a flat band the inverse band effective mass is zero, so the simple single band theory would predict zero superfluid weight $D_s$, i.e., no supercurrent. The Cooper pairs would just be localized and immobile like single particles are in a flat band. Here the geometric part of $D_s$ comes to the rescue: it can be nonzero even in a flat band! Whether it is nonzero turns out to be determined by the band quantum geometry and topology.

In the following, we first give a brief reminder of the basic concepts of quantum geometry and then discuss the concept of superfluid weight. After that the problem of two particles in a flat band is considered: this will give intuition to the many-body results derived later. We treat the many-body problem by utilizing multiband BCS mean-field theory. The superfluid weight is calculated using two approaches and it is shown how the conventional and geometric contributions of superfluidity emerge. The results presented here are based on~\cite{Huhtinen2022,Peotta2022} which remove the caveat related to the orbital-dependence of the results presented in the original works on the quantum geometric superconductivity~\cite{Peotta2015,Liang2017}. We discuss how $D_s$ determines the BKT temperature of superfluidity in two dimensions. In the end, we mention how quantum geometry affects bosonic BECs in a flat band, and discuss systems where quantum geometric superconductivity and superfluidity are relevant and could be experimentally tested, including moir\'e systems~\cite{Torma2022}. 

\section{Basics of quantum geometry}
\label{sec.quantum_metric}


As anticipated in the introduction, for a superconducting state emerging from a partially filled flat band, the transport properties depend in a crucial way on the wave functions of the band. The Bloch theorem states that the eigenfunctions of an electron in a periodic (lattice) potential take the form of Bloch plane waves $\ket{\psi_{n\vc{k}}} = e^{i\vc{k}\cdot\vc{r}}\ket{n_{\vc{k}}}$, where $n$ denotes the band number, $\vc{k}$ the quasimomentum  and $\ket{n_{\vc{k}}}$ is the periodic Bloch function, that is a wave function with the same translational symmetry as the lattice potential.  The periodic Bloch functions can be multiplied by an arbitrary phase factor $\ket{n_{\vc{k}}} \to e^{i\theta(\vc{k})}\ket{n_{\vc{k}}}$, however physical observables never depend on how the phase factor is chosen (sometimes called a gauge choice). Therefore, the Bloch functions can enter the expressions of measurable quantities only through band structure invariants, that is combination of the Bloch functions that do not depend on the phase choice.  

The most important invariant, which will play a crucial role in the following, is the \textit{quantum geometric tensor (QGT)}~\cite{Provost:1980,resta2011}
\begin{equation}\label{abelianQGT}
\begin{split}
\mathcal{B}_{ij}(\vc{k}) & = 2\mathrm{Tr}\big[P(\vc{k})\partial_{k_i}P(\vc{k})\partial_{k_j}P(\vc{k})\big]\,,
\end{split}
\end{equation}
where $P(\vc{k}) = \ket{n_{\vc{k}}}\bra{n_{\vc{k}}}$ is the projector associated with the periodic Bloch function of the $n$-th band. Note that the projector is invariant under the gauge transformation $\ket{n_{\vc{k}}} \to e^{i\theta(\vc{k})}\ket{n_{\vc{k}}}$, therefore also the QGT is invariant. 
The QGT is a Hermitian metric on the specific manifold of quantum states under consideration, in our case the Brillouin zone. In general the QGT can be defined for arbitrary manifolds of quantum states, not just the Brillouin zone. It follows from the definition~\eqref{abelianQGT} that for any $\vc{k}$ the complex matrix $\mathcal{B}_{ij}(\vc{k})$ is positive semidefinite. As a reminder, a positive semidefinite complex matrix $A_{ij}$ has the property that $\sum_{ij}b_i^*A_{ij}b_j \geq 0$ for arbitrary complex numbers $b_i$. This also implies that all its eigenvalues are non-negative.

The real part of the QGT 
\begin{equation}
g_{ij}(\vc{k}) = \mathrm{Re}\,\mathcal{B}_{ij}(\vc{k}) = \mathrm{Tr}[\partial_{k_i}P(\vc{k})\partial_{k_j}P(\vc{k})]
\label{QmetricDefinition}
\end{equation}
is known as the \textit{quantum metric}. The last equality in the above equation follows from Eq.~\eqref{abelianQGT} and the defining property of orthogonal projectors $P(\vc{k}) = P^\dagger(\vc{k}) = P^2(\vc{k})$. The quantum metric
is the infinitesimal form of the Bures distance between pure states. The Bures distance between two normalized pure states $\ket{\psi_1}$ and $\ket{\psi_2}$ is defined as $D_{\rm Bures}(\psi_1,\psi_2) = 2(1-|\langle\psi_1 | \psi_2\rangle|^2)$ in terms of the fidelity  $|\langle\psi_1 | \psi_2\rangle|^2$, which is invariant under multiplication of both states by a phase factor $\ket{\psi_i}\to \ket{\psi'_i} =  e^{i\phi_i}\ket{\psi_i}$. 

On the other hand, the imaginary part of the QGT is the well known Berry curvature, whose surface integral is the Berry geometric phase. It is useful to introduce the Berry curvature starting from the Berry connection
\begin{equation}
 \bm{\mathcal{A}}(\vc{k}) = i \braket{n_{\vc{k}} | \bm{\nabla}_{\vc{k}}n_{\vc{k}} } \,,
\end{equation}
which is not an invariant since it transforms as $\bm{\mathcal{A}}(\vc{k}) \to \bm{\mathcal{A}}(\vc{k})-\bm{\nabla}_{\vc{k}}\theta(\vc{k})$ when the periodic Bloch functions are multiplied by the phase factor $e^{i\theta(\vc{k})}$. Thus the transformation properties of the Berry connection are mathematically the same as the vector potential of electromagnetism, which is the reason why the name \qql gauge\qqr transformation is used even for periodic Bloch functions. The analogy with electromagnetism naturally suggests how to construct band structure invariants from the Berry connection, namely by considering its line integral  along a closed curve $\gamma = \partial S$ at the boundary of a surface $S$ 
\begin{equation}
\begin{split}
\Phi_{\rm Berry} &= \oint_\gamma d\vc{k}\cdot\bm{\mathcal{A}}(\vc{k}) = \int_{S} {\rm d}\vc{S}\cdot \bm{\nabla}_{\vc{k}} \times \mathcal{A}(\vc{k}) \\
&= \frac{1}{2}\int_{S} {\rm d}S_l\varepsilon^{lmn}\mathrm{Im}\,\mathcal{B}_{nm}(\vc{k})\,,
\end{split}
\end{equation}
In the language of electromagnetism, the \textit{Berry phase} $\Phi_{\rm Berry}$ is the magnetic flux through the surface $S$ associated with the magnetic field $\bm{\nabla}_{\vc{k}} \times \bm{\mathcal{A}}(\vc{k})$, which is proportional to the imaginary part of the QGT (the \textit{Berry curvature}).

If the Berry curvature is integrated over the whole Brillouin zone, one obtains the Chern number
\begin{equation}
\mathcal{C} = \frac{1}{2\pi} \int_{\rm BZ}{\rm d}^2\vc{k}\, \mathrm{Im}\, \mathcal{B}_{12}(\vc{k})\,,
\label{eq:Chernnumber}
\end{equation}
a topological invariant of the band structure that is well defined and takes only integer values if the band (or a subset of bands) is isolated.  Since the groundbreaking work of Thouless, Kohmoto, Nightingale and den Nijs~\cite{thouless1982}, it has been known that the Berry curvature and the Chern number play an essential role in the integer quantum Hall effect and more generally in the classification of topological insulators and superconductors~\cite{Hasan2010,bernevig_topological_2013}. 
On the other hand, interesting applications of the quantum metric have begun to emerge only more recently.
Among the first ones, the (Brillouin zone integral of the) quantum metric is encountered as the gauge invariant part of the widely used localization function for Wannier functions introduced by Marzari and Vanderbilt (see Sec.~\ref{sec:Wannier_overlap}). 
Here, we will see that geometrically nontrivial bands, that is bands with nonzero quantum metric, are the only ones that can support superfluid transport in the flat band limit.

If these concepts are new to you and you feel the above description was too brief, we recommend Ref.~\cite{resta2011}.

\section{Superfluid weight}
\label{sec.sf_weight}


	The simplest phenomenological description of the electromagnetic response of superconductive materials is based on the London equations,~\cite{London1935,Tinkham2004} which follow from the constitutive relation 
	\begin{equation}
		\label{eq:London_equation}
		\vc{j} = -D_s\vc{A}\,,
	\end{equation}
	with $\vc{j}$ the current density and $\vc{A}$ the vector potential. The  vector potential is not an invariant quantity under electromagnetic gauge transformation (see previous Sec.~\ref{sec.quantum_metric}), therefore the above relation can be valid only in a specific gauge, called the London gauge.
	
	Equation~\eqref{eq:London_equation} describes a purely non-dissipative response, fundamentally different from the usual normal-state behavior of a material to the application of an external electromagnetic field, which is dissipative and characterized by Ohm's law $\vc{j} =\sigma \vc{E}$, with $\vc{E}= - \partial{\vc{A}}/\partial t$ as the  electric field. The counterpart of the conductivity $\sigma$ for a superconductor is the so-called superfluid weight $D_s$~\cite{Scalapino1992,Scalapino1993}, which is an intrinsic property of the material.  
	Hence a nonzero superfluid weight $D_s\ne 0$ is the very criterion of superconductivity. Equation~\eqref{eq:London_equation} in tandem with Maxwell's equations provides a quantitative description of the two essential phenomena that define the superconductive state, perfect conductivity and perfect diamagnetism. In a perfect diamagnetic material, an external magnetic field is completely screened by the induced magnetization and can penetrate inside the material only up to a characteristic length scale $\lambda_{\rm L}$. This is known as the Meissner effect in the context of superconductivity~\cite{Meissner1933,Tinkham2004}. 
	The penetration depth is essentially the same observable as the superfluid weight since the two are related by  $\lambda_{\rm L} = (\mu_0 D_{\rm s})^{-1/2}$ ($\mu_0$ is the magnetic constant), a simple consequence of Eq.~\eqref{eq:London_equation} together with Amp\`ere's law.
	
	An alternative way to understand the superfluid weight, which provides a practical way to compute it theoretically, is as the energy cost of phase fluctuations away from the equilibrium mean-field configuration of the superconductor~\cite{Gennes1966}.
	Superconductivity is an example of the phenomenon of symmetry breaking in which an order parameter describing the thermodynamic state of the system becomes nonzero. For superconductors the order parameter is a complex field $\Delta(\vc{r}) = |\Delta(\vc{r})| e^{2i\phi(\vc{r})}$. The superfluid weight measures the energy required to create a modulation of the order parameter phase $\phi(\vc{r})$, which is expressed by a term in the free energy of the form~\cite{Weinberg1986}
	\begin{equation}
		\label{eq:free_energy_Ds}
		\Delta F = \frac{\hbar^2}{2e^2}\int {\rm d}^3{\vc{r}}\, \sum_{ij} [D_s]_{ij}\partial_{i}\phi(\vc{r}) \partial_{j}\phi(\vc{r})\,.    
	\end{equation}
	In general the superfluid weight is a rank-2 tensor $[D_s]_{ij}$ in anisotropic systems. Another length scale, the coherence length $\xi_0$,~\cite{Tinkham2004} characterizes the spatial variations of the order parameter amplitude $|\Delta(\vc{r})|$. We also note that Eqs.~\eqref{eq:London_equation} and~\eqref{eq:free_energy_Ds} are not independent relations. The vector potential and the order parameter are not invariant under gauge transformations, but their combination $\bm{\nabla}\phi -e\vc{A}/\hbar$ is an invariant~\cite{Weinberg1986}. It follows that a gradient in the order parameter phase amounts to a finite $\vc{A}$ and hence a finite supercurrent in the system by Eq.~\eqref{eq:London_equation}. Equation~\eqref{eq:free_energy_Ds} can then be interpreted as the kinetic energy associated to the flow of the charge carriers.

	The remarkable and useful properties of superconductors are directly related to the fact that the order parameter phase is very stiff due to Eq.~\eqref{eq:free_energy_Ds} and remains constant or varies slowly over macroscopic length scales. For this reason the superfluid weight is also known as the superfluid stiffness~\cite{Fisher1973}. Thanks to this form of macroscopic phase coherence, quantum mechanical phenomena such as interference and tunneling can occur over time and length scales at which they would usually be destroyed by decoherence. For instance, the interference of the order parameter phase allows SQUIDs (Superconducting QUantum Interference Devices~\cite{Kleiner2004,Tinkham2004}) to measure magnetic fields with exceeding precision. Thermal fluctuations tend to randomize the phase and destroy coherence, and are one of the main limiting factors for SQUIDs. If the thermal fluctuations of the phase are large enough, the material is driven out of the superconducting state and reverts back to the normal resistive state. Thermal fluctuations of the phase are the dominant effect in driving the phase transition  in two dimensional systems~\cite{Berezinskii1971,Berezinskii1972,Kosterlitz1973,Nelson1977} (see Sec.~\ref{sec:BKT_temperature}) and also play an important role in high-$T_{\rm c}$ and Type-II superconductors~\cite{Carlson1999,Emery1995}.
	
	The essential goal of any microscopic theory of superconductivity is to predict phenomenological parameters, in particular the superfluid weight and the coherence length, from first principles. The standard Bardeen-Cooper-Schrieffer (BCS) theory \cite{Bardeen1957,Bardeen1957a,Schrieffer1964,Chandrasekhar1993,Prozorov2006,Aronov1981,Larkin1969,Eilenberger1968,Alexander1985,Choi1989,Kogan2002,Kogan2009,Prozorov2011} gives for the superfluid weight of a spinful electron band at zero temperature the following result
	\begin{equation}
		\label{eq:Ds_BCS_single_band}
		\begin{split}
			[D_s]_{ij} &= \frac{e^2}{\hbar^2} \int \frac{{\rm d}^D\vc{k}}{(2\pi)^D} \, f\big(\epsilon(\vc{k})\big)\frac{\partial^2 \epsilon(\vc{k})}{\partial k_i \partial k_j}  \\
			&=
			\frac{e^2}{\hbar^2} \int \frac{{\rm d}^D\vc{k}}{(2\pi)^D} \, \big[-\partial_\epsilon f\big(\epsilon(\vc{k})\big)\big]  \partial_{k_i}\epsilon(\vc{k}) \partial_{k_j} \epsilon(\vc{k})\,,
		\end{split}
	\end{equation}
	where $D$ is the spatial dimension and $\epsilon(\vc{k})$ the dispersion of the only partially filled band. The quasimomentum integral extends over the whole Brillouin zone (the same convention is used also in the following). 
	The function $f(\epsilon)$ gives the occupation of a single-particle state with energy $\epsilon$ in the BCS ground state.~\cite{Schrieffer1964,Parravicini2013,Tinkham2004} Due to interactions, this is a smoothed version of the Fermi-Dirac distribution $n_F(\epsilon) = \theta(\mu -\epsilon)$ ($\theta(x)$ is the Heaviside step function), even at zero temperature.
	The band dispersion $\epsilon(\vc{k})$ is a single-particle quantity and the most important microscopic property affecting the superfluid weight, according to Eq.~\eqref{eq:Ds_BCS_single_band}. It is evident from the second line of the above equation, that the group velocity 
    \begin{equation*}
    \vc{v}(\vc{k}) = \hbar^{-1}\bm{\nabla}_{\vc{k}}\epsilon(\vc{k})
    \end{equation*}
    close to the Fermi energy plays a crucial role in determining the superfluid weight since the derivative of the occupation function $-\partial_\epsilon f(\epsilon)$ has a peak at the Fermi energy.
	
	By assuming an effective mass approximation for the dispersion $\epsilon({\vc{k}}) \approx \hbar^2\vc{k}^2/(2m_{\rm eff})$, Eq.~\eqref{eq:Ds_BCS_single_band} gives the simple result $D_{\rm s}  = e^2n/m_{\rm eff}$ with $n$ the total number density and $m_{\rm eff}$ the particle effective mass (remember that we are considering the zero temperature case here). This result is independent of the precise form of the occupation function $f(\epsilon)$.
    The physical interpretation is that all the particles take part in the dissipationless flow at zero temperature. This result follows at a general level from continuous translational invariance and time-reversal symmetry~\cite{Leggett1998}. Notably, interactions do not enter  this last result, since the superfluid weight is completely expressed in terms of single-particle quantities: the particle density and the effective mass. The superfluid weight decreases with increasing temperature. The usual interpretation is that only a (superfluid) fraction $0 \leq n_{\rm s}/n \leq 1$ of the charge carriers participate in the superfluid current, the remaining part $1-n_{\rm s}/n$ instead forms a normal dissipative fluid~\cite{London1935,Tinkham2004,Schrieffer1964}. In terms of the superfluid number density $n_{\rm s}$, the superfluid weight then reads 
    \begin{equation*}
     D_{\rm s} = e^2n_{\rm s}/m_{\rm eff}   
    \end{equation*}
    since one may reasonably assume that the effective mass obtained from the single-particle band dispersion does not depend on temperature. The superfluid density vanishes at the critical temperature $T_{\rm c}$ and the material reverts back to its normal state. 
	
	While the superfluid weight is routinely measured in superconducting materials, it is not always clear how to define and measure the superfluid number density $n_{\rm s}$. Therefore it might be misleading to interpret the suppression of $D_{\rm s}$ as a change of the superfluid number density alone and assume that the effective mass is a fixed property which depends only on the single-particle band structure. One goal of these lecture notes is to illustrate a number of recent theoretical results~\cite{Peotta2015,Julku2016,Tovmasyan2016,Liang2017,Huhtinen2022,Torma2018,Tovmasyan2018,Kumar2021,Pyykkonen2021,Iskin2019,Iskin2021twobody,Iskin2022twobody,Peri2021,Herzog-Arbeitman2021,Herzog-Arbeitman2022,Mondaini2018,Chan2021,Chan2022,Kitamura2021,Hofmann2020,Xie2020a,Hu2019,Julku2020,WangLevin2020} pointing out that the \qql effective mass\qqr of the superfluid current carriers is determined not only by the band dispersion, but is also affected by a subtle interplay between interactions and the properties of the single-particle lattice wavefunctions. This is particularly important in the extreme limit of large single-particle effective mass, which would result in a vanishing superfluid weight according to Eq.~\eqref{eq:Ds_BCS_single_band}. In fact  it was pointed out \cite{Basov2011} that perfectly flat \emph{single} electronic bands should  have  zero superfluid weight. To best understand why Eq.~\eqref{eq:Ds_BCS_single_band} is fundamentally incomplete (in multi-band systems), it is useful to study the two-body problem in a flat band~\cite{Torma2018}, that is in a band in which the dispersion $\epsilon(\vc{k})$ is constant and the effective mass is infinite. The solution to this problem is summarized in the next section.

	\section{The two-body problem in a flat band}
	\label{sec:two_body}
	
	The breakthrough in the development of the current theory of the superconducting state came with the work of Leon Cooper who showed that a well-defined Fermi sea is unstable under arbitrarily small attractive interaction between electrons~\cite{Cooper1956,Schrieffer1964}.  In three dimensions the strength of a short-range attractive interaction needs to be larger than a finite value for two-body bound states to occur. This can be traced back to the fact that the density of states of a quadratic band in three dimensions is zero at zero energy, while it is nonzero in one and two dimensions~\cite{Parravicini2013}.  Bound states form for any interaction strength in these latter cases. However, when assuming that the scattering states of the two particles are restricted by Pauli blocking due to a Fermi sea, as Cooper did in his famous Cooper-problem calculation, then bound states occur for any value of the interaction strength even in $d = 3$ since what counts in this case is the nonzero density of states at the Fermi energy.
	
	In crystalline materials, the electron dispersion and the density of states are heavily affected by the atomic lattice~\cite{Parravicini2013}. In particular the electron effective mass can be much larger (or smaller) than its free space value. With increasing effective mass the threshold for the formation of a bound state becomes smaller and smaller. In the flat band limit of infinite effective mass two-body bound states always occur for any small value of the attractive interaction~\cite{Torma2018} - as the density of states at zero energy is no longer vanishing. To show this, one can consider the two-particle scattering problem in the limit where the interaction energy is much smaller than the gaps between the flat and other bands. The two-particle Schr\"odinger equation becomes
	\begin{equation}
	\label{Start2}
	\begin{split}
|\psi (1,2)\rangle =& |\varphi_0(1,2)\rangle + \frac{1}{E-E_0}\sum_{k\neq 0} |\varphi_k\rangle \langle \varphi_k | \lambda V_I| \psi(1,2)\rangle  \\
E_b &\equiv E-E_0 = \langle \varphi_0 | \lambda V_I| \psi(1,2)\rangle  ,
\end{split}
\end{equation}
where $k$ refers to quantum numbers in the isolated flat band. We introduced the notation $E_b$ for the pair binding energy; $\lambda$ is the strength of the interaction and $V_I$ the interaction potential.
	
One can solve this, within the separable potential approximation as the original Cooper problem, and obtain the result
\begin{equation}
E_b = \frac{\lambda}{N_c} \sum_{\mathbf{k}} \int d\mathbf{x} V(\mathbf{x}) | n_{\mathbf{k} + \frac{\mathbf{q}}{2}} (\mathbf{x}) n_{\mathbf{k} - \frac{\mathbf{q}}{2}} (\mathbf{x})|^2 
. \label{NonUnifFinal}
\end{equation}
Here $\mathbf{x}$ is the orbital coordinate and $N_c$ is the number of unit cells. This result shows that indeed there is a bound state for any attractive interaction strength $\lambda$, no matter how small. Moreover, and very importantly, the effective mass is determined by the overlap of Bloch functions differing by the center of mass (pair) momentum $\mathbf{q}$, and depends on $\mathbf{q}$. This hints that the pair could have a dispersion.

Expanding with small $\mathbf{q}$ one can indeed obtain the effective mass. Due to time-reversal symmetry, the point $\mathbf{q}= \vc{0}$ in the Brillouin zone is an extremum of the dispersion. Then from Eq.~\eqref{NonUnifFinal}, we obtain an estimate for the effective mass
	tensor of the bound state
	\begin{equation}
		\label{eq:eff_mass_flat_qm}
		\begin{split}
			\left[\frac{1}{m_{\rm eff}}\right]_{ij}  &\approx \frac{UV_c}{\hbar^2}\int \frac{{\rm d}^D\vc{k}}{(2\pi)^D}\,\sum_{\alpha = 1}^{N_{\rm orb}} \partial_{k_i}P_{\alpha\alpha}(\vc{k})\partial_{k_j}P_{\alpha\alpha}(\vc{k}) \\
			&\approx \frac{UV_c}{N_{\rm orb}\hbar^2}\int \frac{{\rm d}^D\vc{k}}{(2\pi)^D}\,\mathrm{Tr}\big[ \partial_{k_i}P(\vc{k})\partial_{k_j}P(\vc{k})\big]\,.
		\end{split}
	\end{equation}
	Here $V_c$ is the volume of the unit cell, and $N_{\rm orb}$ is the number of orbitals.
	We used the approximation 
	\begin{equation}
	\sum_{\alpha = 1}^{N_{\rm orb}}\partial_{k_i}P_{\alpha\alpha}(\vc{k})\partial_{k_j}P_{\alpha\alpha}(\vc{k}) \approx N_{\rm orb}^{-1}\sum_{\alpha,\beta = 1}^{N_{\rm orb}}\partial_{k_i}P_{\alpha\beta}(\vc{k})\partial_{k_j}P_{\beta\alpha}(\vc{k})
	\end{equation}
	to obtain the second line~\cite{Torma2018}. The quantity under the integral sign in the second line is the quantum metric defined in Eq.~(\ref{QmetricDefinition}).
 
As shown in Ref.~\cite{Torma2018}, the approximation provided by Eq.~\eqref{eq:eff_mass_flat_qm} is usually rather good; if the quantum metric does not have zero eigenvalues then the $\Gamma$ point  $\vc{q} = \vc{0}$ is a minimum of the dispersion. Recent work~\cite{Huhtinen2022,Herzog-Arbeitman2022} has shown that it is actually the minimal quantum metric that determines the two-body bound state effective mass, as will be discussed in Section~\ref{sec.sfw_fb}.
	
	Equation~\eqref{eq:eff_mass_flat_qm} teaches us that the effective mass of a Cooper pair in a flat band is determined by both the interaction strength and the quantum metric, which is purely a property of the single-particle wavefunctions. The physical interpretation of the quantum metric is discussed  in Sec.~\ref{sec:Wannier_overlap}. The flat bands with nonzero quantum metric are physically interesting as they support propagating Cooper pairs and, as  seen below, a robust superconducting state. 
	
	Interestingly (this is one of the remarkable features of isolated flat bands), one can derive the same result for the pair effective mass by starting from a projected many-body Hamiltonian instead of the two-body problem. To illustrate this point concretely, consider spin-$1/2$ fermions moving in a lattice  and interacting by means of an attractive  Hubbard interaction of the form
 \begin{equation}
 H_{\rm int} = U\sum_{\vc{i}\alpha} n_{\vc{i}\alpha\uparrow}n_{\vc{i}\alpha\downarrow} .
 \end{equation}
 Here $U<0$, and $n_{\vc{i}\alpha\sigma} = \crea{c_{\vc{i}\alpha\sigma}}\ani{c_{\vc{i}\alpha\sigma}}$
 is the occupation number operator of the lattice site labelled by the unit cell index $\vc{i} = (i_1,i_2,i_3)$, which is a triplet of integers in dimension $D = 3$, and the sublattice index $\alpha$, while $\sigma =\uparrow,\downarrow$ is the spin index. As emphasized by this notation we consider the general case of a composite lattice with orbitals $\alpha$, that is, a lattice composed of several simple (Bravais) lattices labelled by $\alpha$. 
	For instance the $s$ or $p_z$ orbitals on the hexagonal lattice of graphene, which is composed of two triangular lattices, have the sublattice index taking two values $\alpha = 1,2$.

	In the isolated band limit only a single energy band (per spin) of the lattice model plays a significant role since all other bands are separated by a band gap $E_{\rm gap}$  larger than the Hubbard coupling $|U|$. In this limit it becomes a good approximation to restrict the Hilbert space to the subspace spanned by the states in the chosen band.~\cite{Huber2010,Tovmasyan2013,Tovmasyan2016,Torma2018} Furthermore, if the band is exactly flat, the only nontrivial part of the projected Hamiltonian is $\overline{H} = P_n H_{\rm int} P_n$, where $P_n$ is the projection operator on the relevant band, labelled by $n$ (for a precise definition of the projection operator see Ref.~\cite{Tovmasyan2016}). After the projection on the flat band, the problem is still genuinely a many-body one: $\overline{H}$  is a sum over squares of projected density operators which are rendered non-commuting due to the projection operation. Nevertheless, under the assumption of time-reversal symmetry, spin rotation symmetry around a given axis (for convenience the $z$-axis) and a condition on the flat band wavefunctions called the uniform pairing condition, one can show that the state
	\begin{equation}
	\label{eq:two-body_bound_state}
	\ket{\Psi_{0}} = \sum_{\vc{k}}f^\dagger_{\vc{k}\uparrow}f^\dagger_{-\vc{k}\downarrow}\ket{0}
	\end{equation}
	is an exact eigenstate of $\overline{H}$  with energy $E_0 = -|U|/N_{\rm orb}$~\cite{Tovmasyan2016}. (The uniform pairing condition means the superconducting order parameter is the same in all orbitals where it is nonzero; this is the case in general for sufficiently symmetric systems~\cite{Herzog-Arbeitman2022}.) Here, by $f_{\vc{k}\sigma},\,f_{\vc{k}\sigma}^\dagger$ we denote the fermionic flat band eigenstate operators of the non-interacting Hamiltonian related to the physical electron by the projection 
    \begin{equation*}
    P_n c_{\vc{i}\alpha}P_n = \sum_{\vc{k}} \psi_{n\vc{k}\sigma}(\vc{i},\alpha)f_{\vc{k}\sigma}.        
    \end{equation*}
    The eigenstates of the non-interacting Hamiltonian take the form of Bloch plane waves
    \begin{equation*}
    \psi_{n\vc{k}\sigma}(\vc{i}\alpha) = e^{i\vc{k}\cdot\vc{r}_{\vc{i}\alpha}}n_{\vc{k}\sigma}(\alpha)/\sqrt{N_{c}}        
    \end{equation*}
    ($\vc{r}_{\vc{i}\alpha}$ is the position vector of the lattice site labelled by $\vc{i}\alpha$). The parameter $N_{\rm orb}$, appearing in the energy eigenvalue $E_0$, is the number of orbitals on which the flat band wavefunctions have nonzero amplitude, given by the number of distinct values of $\alpha$ for which $\int {\rm d}^D\vc{k} \,|n_{\vc{k}\uparrow}(\alpha)|^2 \neq 0$~\cite{Peotta2015,Julku2016,Tovmasyan2016}.
	
	The state $\ket{\Psi_0}$ is a two-particle bound state: the wavefunction amplitude decays to zero with increasing distance between the two electrons. Moreover it  can be also shown that there are no other two-particle states with energy lower than $E_0$. Interestingly, this exact solution can be straightforwardly generalized to an arbitrary number of pairs of particles~\cite{Tovmasyan2016}. Crucially, in a flat band a bound state always forms under the above conditions, even in three dimensions. While superficially this is the same result obtained by solving the Cooper problem, there are some essential differences. In the case of a flat band there is no need to consider a Fermi sea since the density of states is nonzero to begin with, and in fact diverges in a flat band. The concept of Fermi sea itself is not well-defined in a flat band, where all the states have the same energy. Moreover, the bound state energy is linearly proportional to the coupling constant $U$, whereas in the original Cooper problem it has a non-analytic dependence $E_0 \propto \exp\big(-1/|U|\rho(\varepsilon_{\rm F})\big)$, where $\rho(\varepsilon_{\rm F})$ is the density of states at the Fermi energy.
	
A crucial question is: what is the effective mass of the bound state described by $\ket{\Psi_0}$? Naively, one could expect that since the constituent particles have infinite effective mass this would be the case also for the bound state they form. This would mean that the superfluid weight is zero and no superconducting state can occur in a partially filled flat band. However, there is no reason why a finite momentum deviation of $\ket{\Psi_0}$ cannot disperse, and hence the bound-state mass can, generically, be nonzero. An estimate of the dispersion of the two-body bound state is obtained from~\cite{Torma2018}
\begin{equation}
		\label{eq:Eq}
		\begin{split}
			E(\vc{q}) &= \mathrm{Tr}_{2,\vc{q}}[\overline{H}] = \sum_{\vc{k}}\bra{0}f_{\vc{q}-\vc{k},\downarrow}f_{\vc{k}\uparrow}
			\overline{H}
			f_{\vc{k}\uparrow}^\dagger f_{\vc{q}-\vc{k},\downarrow}^\dagger\ket{0} \\
			&= -UV_{\rm c}\int \frac{{\rm d}^D\vc{k}}{(2\pi)^D}\,\sum_{\alpha = 1}^{N_{\rm orb}} P_{\alpha\alpha}(\vc{k})P_{\alpha\alpha}(\vc{k}-\vc{q})\,.
		\end{split}
	\end{equation}
In the above equation we use the fact that the Hamiltonian $\overline{H}$ has two conserved quantities, the particle number and the quasimomentum, and we indicate with $\mathrm{Tr}_{2,\vc{q}}[\overline{H}]$ the trace over the Hamiltonian block corresponding to particle number $N=2$ and quasimomentum $\vc{q}$. This trace is defined explicitly on the right-hand side of the first line of Eq.~\eqref{eq:Eq}. In the second line, $V_{\rm c}$ is the unit cell volume and we denote by $P(\vc{k})$ the projector on the Bloch state with momentum $\vc{k}$, whose matrix elements are
\begin{equation*}
 P_{\alpha\beta}(\vc{k}) = n_{\vc{k}\uparrow}(\alpha)n^*_{\vc{k}\uparrow}(\beta).   
\end{equation*}
Only the spin-$\uparrow$ Bloch functions appear since the time-reversal symmetry  constrain $n_{\vc{k}\uparrow}(\alpha) = n^*_{-\vc{k},\downarrow}(\alpha)$ has been used. By the general property of the trace, $E(\vc{q}) = \sum_m E_m(\vc{q})$ is the sum of all the eigenvalues of $\overline{H}$ corresponding to states with $N = 2$ particles and quasimomentum $\vc{q}$. In particular $E_0(\vc{q})$  is the dispersion of the lowest bound state with energy $E_0 = E_0(\vc{q}=\vc{0})$, as presented above. 
	
Most examples of realistic lattice models with perfectly flat bands exhibit a dispersing  $E(\vc{q})$ (Eq.~\eqref{eq:Eq}) as a function of quasimomentum. Hence there is at least one two-particle state which possesses a non-flat dispersion and is thus propagating. This is necessarily a bound state since particles that are far apart do not interact. 
In practice, it is observed that the lowest bound state is the one with the largest bandwidth, thus $E(\vc{q})$ is a good approximation of $E_0(\vc{q})$ (this can be further justified using a separable potential approximation~\cite{Torma2018}). In fact, it is easy to show that there are at most $N_{\rm orb}$ propagating bound states; this is usually a small number - the number of orbitals where the flat band states have a nonzero weight. Furthermore, if there is only a single propagating bound state then Eq.~\eqref{eq:Eq} gives its dispersion exactly up to a constant term.
	
From Eq.~\eqref{eq:Eq}, we obtain an estimate for the effective mass tensor of the bound state (using the approximations mentioned above Eq.~\eqref{eq:two-body_bound_state})
\begin{equation}
		\label{eq:eff_mass_flat}
		\begin{split}
			\left[\frac{1}{m_{\rm eff}}\right]_{ij}  &\approx \frac{UV_c}{\hbar^2}\int \frac{{\rm d}^D\vc{k}}{(2\pi)^D}\,\sum_{\alpha = 1}^{N_{\rm orb}} \partial_{k_i}P_{\alpha\alpha}(\vc{k})\partial_{k_j}P_{\alpha\alpha}(\vc{k}) \\
			&\approx \frac{UV_c}{N_{\rm orb}\hbar^2}\int \frac{{\rm d}^D\vc{k}}{(2\pi)^D}\,\mathrm{Tr}\big[ \partial_{k_i}P(\vc{k})\partial_{k_j}P(\vc{k})\big]\,.
		\end{split}
	\end{equation}
This is the same result as Equation~(\ref{eq:eff_mass_flat_qm}).
	
A final important remark is that in order to have Cooper pairs with finite effective mass it is necessary that $N_{\rm orb} > 1$. This can be seen from Eq.~\eqref{eq:eff_mass_flat} together with the normalization property $\mathrm{Tr}\left[P(\vc{k})\right] = 1$. For a single band $N_{\rm orb} = 1$ this implies $P(\vc{k})=1$  and the problem of diagonalizing the projected Hamiltonian becomes trivial. This means that Cooper pair transport in a flat band and ---as a consequence--- flat band superconductivity, is a new qualitative effect that can occur only in the context of genuinely multiband/multiorbital lattices with strong inter-orbital mixing. 
	
These results show that Cooper pairs can have a finite effective mass and thus support transport if the flat band quantum metric is nonzero. This is in sharp contrast with Eq.~\eqref{eq:Ds_BCS_single_band}, which gives zero superfluid weight for a partially filled single (or non-hybridizing) flat band, showing that Eq.~\eqref{eq:Ds_BCS_single_band}
neglects additional contributions that come into play in the case of multiband/multiorbital lattices ($N_{\rm orb} > 1$). These were identified only recently by applying standard BCS theory to multiband lattice models~\cite{Peotta2015,Julku2016,Liang2017,Huhtinen2022} and are the subject of the next Section.
	
\section{Conventional and geometric contributions of the superfluid weight}
\label{sec.conv_geom}

\subsection{Superfluid weight of the multiband Hubbard model}
\label{sec.mf_sfw}

In this section, we will derive expressions for the superfluid weight in a generic lattice model, with a Hubbard Hamiltonian
\begin{align}
  H &= H_0 + H_{\rm int}, \label{eq.ham} \\
  H_0 &= \sum_{\sigma}\sum_{i\alpha,j\beta} t_{i\alpha,j\beta}^{\sigma}\crea{c_{i\alpha\sigma}}\ani{c_{j\beta\sigma}}
  -\mu N \label{eq.H0} \\
  H_{\rm int} &= U\sum_{i\alpha} \crea{c_{i\alpha\up}} \crea{c_{i\alpha\dn}}
  \ani{c_{i\alpha\dn}} \ani{c_{i\alpha\up}} \label{eq.Hint_exact}.
\end{align}
We label the unit cells with $i,j$ and the orbitals within a unit cell by $\alpha,\beta$. Particles with spin $\sigma$ can hop from site $j\beta$ to site $i\alpha$ with amplitude $t_{i\alpha,j\beta}^{\sigma}$, and particles on the same site experience an interaction of strength $U$, which we take attractive $U<0$. The chemical potential $\mu$ tunes the particle number $N=\sum_{\sigma,i\alpha}n_{i\alpha}$. Although simple, the Hubbard model is not generally exactly solvable. In the following, we will be using the mean-field approximation
\begin{equation}
H_{\rm int} \approx
\sum_{i\alpha}\left(\Delta_{i\alpha}\crea{c_{i\alpha\up}}\crea{c_{i\alpha\dn}} + \Delta_{i\alpha}^*\ani{c_{i\alpha\dn}}\ani{c_{i\alpha\up}} - |\Delta_{i\alpha}|^2/U\right), \label{eq.Hint_mf}
\end{equation}
where $\Delta_{i\alpha} =
U\ave{c_{i\alpha\dn} c_{i\alpha\up}}$.

As explained in Sec.~\ref{sec.sf_weight}, the superfluid weight measures the difference in free energy $F=\Omega+\mu N$, where $\Omega$ is the grand potential, due to a modulation of the phase of the order parameter. More precisely~\cite{Taylor2006,Peotta2015},
\begin{equation}
  [D_s]_{ij} = \frac{e^2}{V\hbar^2}\frac{{\rm d}^2F}{{\rm d}q_i{\rm
      d}q_j}\bigg|_{\vec{q}=\vec{0}}, \label{eq.sfw}
\end{equation}
where the vector $\vec{q}$ controls the phase of the order parameters though the transformation $\Delta_{i\alpha}\to\Delta_{i\alpha}e^{2i\vec{q}\cdot\vec{r}_{i\alpha}}$, with $\vec{r}_{i\alpha}$ being the position of site $i\alpha$. The system volume is $V=N_cV_c$, where $N_c$ is the number of unit cells. The derivatives ${\rm d}/{\rm d}q_i$ are total derivatives: the implicit dependence of all variables on $\vec{q}$ needs to be taken into account when computing them~\cite{Huhtinen2022}.

We now introduce $\vec{q}$ into the mean-field Hamiltonian, and use the Fourier transformation
\begin{equation}
c_{i\alpha\sigma} =
\frac{1}{\sqrt{N_c}} \sum_{\kk}
e^{i\kk\cdot\vec{r}_{i\alpha}} c_{\kk\alpha\sigma}, \label{eq.fourier}
\end{equation}
to express $H$ in reciprocal space:
\begin{equation}
\label{eq.k_hamiltonian}
\begin{split}
    H(\vec{q}) &= \sum_{\sigma}\sum_{i\alpha,j\beta}t_{i\alpha,j\beta}^{\sigma}
    \crea{c_{i\alpha\sigma}}\ani{c_{j\beta\sigma}} - \mu N +\sum_{i\alpha}\Delta_{i\alpha}e^{2i\vec{q}\cdot\vec{r}_{i\alpha}}
    \crea{c_{i\alpha\up}}\crea{c_{i\alpha\dn}} + {\rm H.c.} - \frac{|\Delta_{i\alpha}|^2}{U} \\
    &= \frac{1}{N_c}\sum_{\sigma}\sum_{i\alpha,j\beta}\sum_{\kk\kk'} t_{i\alpha,j\beta}e^{-i\vec{k}\cdot(\vec{R}_i+\vec{\delta}_{\alpha}-\vec{R}_j-\vec{\delta}_{\beta})}
    e^{i(\kk'-\kk)\cdot(\vec{R}_j-\vec{\delta}_{\beta})} \crea{c_{\kk\alpha\sigma}}\ani{c_{\kk'\beta\sigma}} \\
    &+ \sum_{i\alpha}\sum_{\kk\kk'} 
    \frac{1}{N_c}\Delta_{i\alpha}e^{i(2\qq-\kk-\kk')\cdot\vec{r}_{i\alpha}}\crea{c_{\kk\alpha\up}}\crea{c_{\kk'\alpha\dn}} + {\rm H.c.} - \frac{|\Delta_{i\alpha}|^2}{U} \\
    &= \sum_{\sigma}\sum_{\kk}\sum_{\alpha\beta} [H_{\kk}^{\sigma}]_{\alpha\beta} \crea{c_{\kk\alpha\sigma}}\ani{c_{\kk\beta\sigma}} 
    + \sum_{\kk}\sum_{\alpha\beta} \Delta_{i\alpha}\crea{c_{\qq+\kk,\alpha,\up}}\crea{c_{\qq-\kk,\alpha\dn}} + {\rm H.c.} - \frac{|\Delta_{i\alpha}|^2}{U}. 
    \end{split}
\end{equation}  
Here, 
\begin{equation*}
[H_{\vec{k}}^{\sigma}]_{\alpha\beta} = \sum_{i}
t_{i\alpha,0\beta}^{\sigma} e^{-i\vec{k}\cdot
  (\vec{R}_i+\vec{\delta}_{\alpha}-\vec{\delta}_{\beta})}    
\end{equation*}
is the Fourier transformation of the kinetic Hamiltonian for spin $\sigma$, where $\vec{R}_i$ the position of unit cell $i$, and 
\begin{equation*}
\vec{\delta}_{\alpha} = \vec{r}_{i\alpha}-\vec{R}_i    
\end{equation*}
is the position of the orbital $\alpha$ within a unit cell. To obtain the final result from the second line, we have used the translational invariance of the lattice, which means that
\begin{equation*}
 t_{i\alpha,j\beta}^{\sigma}=t^{\sigma}_{m\alpha,0\beta}   
\end{equation*}
when $\vec{R}_m=\vec{R}_i-\vec{R}_j$.  The Hamiltonian~\eqref{eq.k_hamiltonian} can be rewritten in matrix form as follows:
\begin{equation}
\begin{split}
  &H(\vec{q}) = \sum_{\vec{k}} \vec{\crea{c_{\vec{k}}}}
  H_{\rm BdG}(\vec{q},\vec{k}) \vec{\ani{c_{\vec{k}}}}  
  + \sum_{\vec{k}}{\rm
    Tr} H_{\vec{k}}^{\dn} - N_b N_c\mu - N_c\sum_{\alpha}
  \frac{|\Delta_{\alpha}(\vec{q})|^2}{U}, \\
  &H_{\rm BdG}(\vec{q},\vec{k}) = \begin{pmatrix}
    H_{\vec{q}+\vec{k}}^{\up} - \mu \mathbb{1}_{N_b} & \vec{\Delta} \\
    \vec{\Delta}^{\dag} & - (H_{\vec{q}-\vec{k}}^{\dn})^* + \mu \mathbb{1}_{N_b}
  \end{pmatrix},
  \end{split}
\end{equation}
where
\begin{equation*}
 \vec{\ani{c_{\vec{k}}}} = (
\ani{c_{\vec{q}+\vec{k},\alpha=1,\up}},\ldots,
\ani{c_{\vec{q}+\vec{k},\alpha=N_b,\up}},  
\crea{c_{\vec{q}-\vec{k},\alpha=1,\dn}},\ldots,\crea{c_{\vec{q}-\vec{k},\alpha=N_b,\dn}})^{\rm  
  T} , 
\end{equation*}
$N_b$ is the number of bands per spin, $\mathbb{1}_n$ is the $n\times n$ identity matrix and 
\begin{equation*}
 \vec{\Delta}={\rm diag}(\Delta_{1}(\vec{q}),\ldots,\Delta_{N_b}(\vec{q})).   
\end{equation*}
Note that the order parameters depend on $\vec{q}$: they need to be solved for each value of 
$\vec{q}$. This can be done by minimizing the grand potential
\begin{equation}
  \Omega = -\frac{1}{\beta} \sum_{\vec{k}}\sum_{i} \ln
         [1+\exp(-\beta E_{\vec{q},\vec{k},i})]  
         +\sum_{\vec{k}}{\rm
           Tr}H_{\vec{k}}^{\dn} - N_bN_{c}\mu - N_c\sum_{\alpha}
         \frac{|\Delta_{\alpha}|^2}{U} , 
\end{equation}
where $E_{\vec{q},\vec{k},i}$ are the eigenvalues of the Bogoliubov-de-Gennes
Hamiltonian $H_{\rm BdG}(\vec{q},\vec{k})$ and $\beta=1/T$ is the inverse temperature (in units where $k_B=1$). We thus require that $\partial\Omega/\partial\Delta_{\alpha}=0$ for 
all $\Delta_{\alpha}$ at any $\vec{q}$. The overall phase of the order parameters is free, i.e. we can add the same complex phase to all order parameters without changing the state of the system. In the following, we will fix this overall phase by requiring that one nonzero order parameter is positive and real: we will choose it to be $\Delta_1$ without loss of generality. 
The particle number is $N=-\partial\Omega/\partial \mu$. When computing the 
superfluid weight from Eq.~\eqref{eq.sfw}, we will assume that $\mu$ varies with $\vec{q}$ in such a way that $N$ is kept
constant. 

Since $\Delta_{\alpha}$ generally depends on $\vec{q}$, computing the superfluid weight directly from Eq.~\eqref{eq.sfw} 
requires solving the order parameters at several nonzero $\vec{q}$. Especially for complex lattices with a large number of bands, 
this can be quite a heavy operation. We will now derive an expression for $D_s$ which requires only knowledge of the state at 
$\vec{q}=0$ by expressing the {\it total} derivative ${\rm d}^2F/{\rm d}q_i{\rm d}q_j$ in terms of {\it partial} derivatives 
of $\Omega$.

By applying the chain rule, we can write the first derivative of $\Omega$ as
\begin{equation}
  \frac{{\rm d}\Omega}{{\rm d}q_i} = \frac{\partial \Omega}{\partial
    q_i} + \frac{\partial \Omega}{\partial 
    \mu} \frac{{\rm d}\mu}{{\rm d}q_i} + \sum_{\alpha} \frac{\partial
    \Omega}{\partial \Delta_{\alpha}^{I}} \frac{{\rm
      d}\Delta_{\alpha}^{I}}{{\rm 
      d}q_i} + \sum_{\alpha} \frac{\partial \Omega}{\partial
    \Delta_{\alpha}^{R}} \frac{{\rm d}\Delta_{\alpha}^{R}}{{\rm
      d}q_i},
  \label{eq.step_chain}
\end{equation}
where $\Delta_{\alpha}^I={\rm Im}(\Delta_{\alpha})$ and $\Delta_{\alpha}^R={\rm Re}(\Delta_{\alpha})$. The {\it partial} 
derivative $\partial/\partial q_i$ is taken by varying $q_i$ but keeping all other variables constant, and is different 
from the {\it total} derivative ${\rm d}/{\rm d}q_i$, which requires us to take the dependence of $\mu$ and 
$\Delta_{\alpha}$ on $\vec{q}$ into account. As $\partial \Omega/\partial\Delta_{\alpha}^I=\partial \Omega/\partial \Delta_{\alpha}^R=0$ for all $\vec{q}$ and $N=-\partial \Omega/\partial \mu$ is constant as a function of $\vec{q}$, we obtain
\begin{equation}
\begin{split}
\frac{{\rm d}^2F}{{\rm d}q_i{\rm d}q_j} &= 
\frac{{\rm d}}{{\rm d}q_j}\frac{\partial \Omega}{\partial q_i} + \frac{\partial \Omega}{\partial \mu}\frac{{\rm d}^2\Omega}{{\rm d}q_i{\rm d}q_j} + N\frac{{\rm d}^2\mu}{{\rm d}q_i{\rm d}q_j} \\
&= \frac{\partial^2\Omega}{\partial q_i\partial q_j} +
  \frac{\partial^2\Omega}{\partial \mu\partial q_i} \frac{{\rm
      d}\mu}{{\rm d}q_j} 
  + \sum_{\alpha} \left(
  \frac{\partial^2\Omega}{\partial \Delta_{\alpha}^R\partial q_i}
  \frac{{\rm d}\Delta_{\alpha}^R}{{\rm d}q_j}
  + \frac{\partial^2\Omega}{\partial \Delta_{\alpha}^I\partial q_i}
  \frac{{\rm d}\Delta_{\alpha}^I}{{\rm d}q_j} 
  \right).
  \end{split}
\end{equation}

This equation can be written in a more compact form by using again that the particle number is kept fixed and $\partial \Omega/\partial \Delta_{\alpha}=0$, which implies that
\begin{equation}
  \frac{{\rm d}}{{\rm d}q_i} \frac{\partial \Omega}{\partial
    \Delta_{\alpha}^R} = \frac{{\rm d}}{{\rm d}q_i} \frac{\partial
    \Omega}{\partial 
    \Delta_{\alpha}^I} = \frac{{\rm d}}{{\rm d}q_i} \frac{\partial
    \Omega}{\partial \mu} = 0, 
\end{equation}
which is equivalent to the system of equations
\begin{equation}
(\partial_{\Delta,\mu}^2\Omega)\vec{f}_i = -\vec{a}_i,\label{eq.sys_del_mu}
\end{equation} 
where
\begin{align}
  \partial_{\Delta,\mu}^2\Omega &= \begin{pmatrix}
    \frac{\partial^2 \Omega}{(\partial \Delta_1^R)^2} & \ldots &
    \frac{\partial^2 \Omega}{\partial \Delta_1^R\partial \Delta_{N_b}^R} &
    \frac{\partial^2\Omega}{ \partial \Delta_1^R\partial \Delta_2^I} &
    \ldots &
    \frac{\partial^2\Omega}{\partial\Delta_1^R\partial\Delta_{N_b}^I} &
    \frac{\partial^2\Omega}{\partial \Delta_1^R\partial\mu} \\
    \vdots & \ddots & \vdots & \vdots & \ddots & \vdots & \vdots \\
    \frac{\partial^2\Omega}{\partial \Delta_{N_b}^R\partial \Delta_1^R} &
    \ldots & \frac{\partial^2\Omega}{(\partial \Delta_{N_b}^R)^2} &
    \frac{\partial^2\Omega}{\partial \Delta_{N_b}^R\partial \Delta_2^I} &
    \ldots & \frac{\partial^2\Omega}{\partial
      \Delta_{N_b}^R\partial\Delta_{N_b}^I} & \frac{\partial^2\Omega}{\partial
      \Delta_{N_b}^R\partial \mu} \\
    \frac{\partial^2\Omega}{\partial \Delta_2^I\partial \Delta_1^R} &
    \ldots & \frac{\partial^2\Omega}{\partial \Delta_2^I\partial
      \Delta_{N_b}^R} & \frac{\partial^2\Omega}{(\partial \Delta_2^I)^2} &
    \ldots & \frac{\partial^2\Omega}{\partial \Delta_2^I\partial
      \Delta_{N_b}^I} & \frac{\partial^2\Omega}{\partial
      \Delta_2^I\partial \mu} \\
    \vdots & \ddots & \vdots & \vdots & \ddots & \vdots & \vdots \\
    \frac{\partial^2\Omega}{\partial \Delta_{N_b}^I\partial \Delta_1^R} &
    \ldots & \frac{\partial^2\Omega}{\partial \Delta_{N_b}^I\partial
      \Delta_{N_b}^R} & \frac{\partial^2\Omega}{\partial \Delta_{N_b}^I\partial
      \Delta_2^I} & \ldots & \frac{\partial^2\Omega}{(\partial
      \Delta_{N_b}^I)^2} & \frac{\partial^2\Omega}{\partial
      \Delta_{N_b}^I\partial \mu} \\
    \frac{\partial^2\Omega}{\partial \mu\partial \Delta_1^R} & \ldots &
    \frac{\partial^2\Omega}{\partial \mu\partial \Delta_{N_b}^R} &
    \frac{\partial^2\Omega}{\partial \mu\partial \Delta_2^I} & \ldots
    & \frac{\partial^2\Omega}{\partial \mu\partial \Delta_{N_b}^I} &
    \frac{\partial^2\Omega}{\partial \mu^2}
  \end{pmatrix}, \label{eq.hess} \\
  \vec{f}_i &= \left( \frac{{\rm d}\Delta_1^R}{{\rm d}q_i}, \ldots,
  \frac{{\rm d}\Delta_{N_b}^R}{{\rm d}q_i}, \frac{{\rm d}\Delta_2^I}{{\rm
      d}q_i}, \ldots, \frac{{\rm d}\Delta_{N_b}^I}{{\rm d}q_i},
  \frac{{\rm d}\mu}{{\rm d}q_i} \right)^{\rm T} \label{eq.f} \\
  \vec{a}_i &= \left( \frac{\partial^2\Omega}{\partial q_i\partial
    \Delta_1^R}, \ldots, \frac{\partial^2\Omega}{\partial q_i\partial
    \Delta_{N_b}^R}, \frac{\partial^2\Omega}{\partial q_i\partial
    \Delta_2^I}, \ldots, \frac{\partial^2\Omega}{\partial q_i\partial
    \Delta_{N_b}^I}, \frac{\partial^2\Omega}{\partial q_i\partial \mu}
  \right)^{\rm T}. \label{eq.b} 
\end{align}
The imaginary part of $\Delta_1$ is absent in Eqs.~\eqref{eq.hess} to~\eqref{eq.b} as we required it to be real and positive. The superfluid weight given by Eq.~\eqref{eq.sfw} can thus be written as
\begin{equation}
  \frac{V\hbar^2}{e^2}[D_s]_{ij} = \frac{\partial^2\Omega}{\partial q_i\partial q_j}
  \bigg|_{\vec{q}=\vec{0}} - \vec{f}_i^{\rm T} (\partial_{\Delta,\mu}^2\Omega) \vec{f}_i\big|_{\vec{q}=\vec{0}}.
  \label{eq.sfw_full}
\end{equation}
Both $\partial^2\Omega/\partial q_i\partial q_j\big|_{\vec{q}=\vec{0}}$ and $\partial^2_{\Delta,\mu}\Omega$ 
can be computed with knowledge of only the state at $\vec{q}=\vec{0}$, as they involve only partial derivatives, 
but $\vec{f}_i$ still contains the derivatives of the order parameters and chemical potential with reference to
$\vec{q}$. However, using Eq.~\eqref{eq.sys_del_mu}, when $\partial^2_{\Delta,\mu}\Omega$ is invertible, we get 
\begin{equation*}
\vec{f}_i = -(\partial^2_{\Delta,\mu}\Omega)^{-1}\vec{a}_i,     
\end{equation*}
where $\vec{a}_i$ involves only derivatives of the order parameters. 
If we had not fixed the overall phase of the order parameters, $\partial^2_{\Delta,\mu}\Omega$ would contain a row and a column
with terms involving $\Delta_1^I$, and would be singular: the system of equations Eq.~\eqref{eq.sys_del_mu} would have 
an infinite number of solutions due to the freedom in the overall phase of the order parameters. However, requiring that 
one nonzero order parameter is always real and positive generally guarantees that $\partial^2_{\Delta,\mu}\Omega$ is 
invertible. Thus
\begin{equation}
  \frac{V\hbar^2}{e^2}[D_s]_{ij} = \frac{\partial^2\Omega}{\partial q_i\partial q_j}
  \bigg|_{\vec{q}=\vec{0}} - \vec{a}_i^{\rm T} (\partial_{\Delta,\mu}^2\Omega)^{-1} \vec{a}_i\big|_{\vec{q}=\vec{0}}, \label{eq.sfw_full2}
\end{equation}  
which involves only partial derivatives, and can be computed without solving the state at nonzero $\vec{q}$.

The full superfluid weight given in Eq.~\eqref{eq.sfw_full2} can often be simplified by taking advantage of the 
symmetries of the system. For example, if ${\rm d}\mu/{\rm d}q_i\big|_{\vec{q}=\vec{0}}$ is guaranteed, all terms involving $\mu$ in $\vec{a}_i^{\rm T} (\partial_{\Delta,\mu}^2\Omega)^{-1} \vec{a}_i$ can be ignored. A particularly useful case is systems with time-reversal symmetry (where $H_{\kk}^{\up}=(H_{-\kk}^{\dn})^*$), in which both ${\rm d}\mu/{\rm d}q_i\big|_{\vec{q}=\vec{0}}=0$ and ${\rm d}\Delta_{\alpha}^R/{\rm d}q_i\big|_{\vec{q}=\vec{0}}=0$ are guaranteed (for a proof, see Ref.~\cite{Peotta2015}). Then the superfluid weight simplifies to
\begin{equation}
\label{eq.sfw_trs}
\begin{split}
 \frac{V\hbar^2}{e^2}[D_s]_{ij} &=
  \frac{\partial^2\Omega}{\partial q_i\partial
    q_j}\bigg|_{\vec{q}=\vec{0}} - ({\rm 
    d}_i\Delta^I)^{\rm T} \partial_{\Delta^{I}}^2\Omega ({\rm
    d}_j\Delta^I)\big|_{\vec{q}=\vec{0}},  \\ 
  {\rm d}_i\Delta^I &= \left( \frac{{\rm d}\Delta_2^I}{{\rm d}q_i},
  \ldots, \frac{{\rm d}\Delta_{N_b}^I}{{\rm d} q_i}
  \right)^{\rm T},\\
  \partial_{\Delta^{I}}^2\Omega &= \begin{pmatrix}
    \frac{\partial^2\Omega}{\partial \Delta_{2}^I\partial
      \Delta_{2}^I} & \ldots & \frac{\partial^2\Omega}{\partial
      \Delta_{2}^I\partial \Delta_{N_b}^I} \\
    \vdots & \ddots & \vdots \\
    \frac{\partial^2\Omega}{\partial \Delta_{N_b}^I\partial \Delta_2^I}
    & \ldots & \frac{\partial^2\Omega}{\partial \Delta_{N_b}^I\partial
      \Delta_{N_b}^I} 
  \end{pmatrix},
  \end{split}
\end{equation}
where
\begin{equation}
\label{eq.lin_sys_del}
\begin{split}
  &(\partial_{\Delta^I}^2\Omega){\rm d}_i\Delta^{I} = -\vec{b}_i, \qquad\vec{b}_i = \left( \frac{\partial^2\Omega}{\partial q_i\partial
    \Delta_{2}^{I}},\ldots, \frac{\partial^2\Omega}{\partial
    q_i\partial \Delta_{N_b}^{I}} \right)^{\rm T}.
\end{split}
\end{equation}

The superfluid weight in systems with time-reversal symmetry (TRS) simplifies to 
\begin{equation*}
 [D_s]_{ij}=(e^2/V\hbar^2)\partial^2\Omega/\partial q_i\partial q_j\big|_{\vec{q}=\vec{0}}   
\end{equation*}
when 
\begin{equation*}
 {\rm d}\Delta_{\alpha}^I/{\rm d}q_i\big|_{\vec{q}=\vec{0}}=0   
\end{equation*}
for all $\alpha$ and $i$. This means that in single-band models, the aforementioned simplified equation can always be used: we can make the single order parameter always real. However, in multiband systems, this is not always the case, because the order parameters at nonzero $\vec{q}$ can differ by an orbital-dependent phase, which cannot be removed with a transformation of the overall phase of the order parameters. In such a case, it is essential to also take the terms $({\rm d}_i\Delta^I)^{\rm T}\partial^2_{\Delta^I}\Omega({\rm d}_j\Delta^I)$ in Eq.~\eqref{eq.sfw_trs} into account; otherwise, the result can be quantitatively or even qualitatively wrong. In fact, in systems with TRS, 
\begin{equation}
    \frac{\partial^2\Omega}{\partial q_i^2}\geq \frac{{\rm d}^2F}{{\rm d}q_i^2}. \label{eq.inequality}
\end{equation}
This is easily seen by noticing that $\partial^2_{\Delta^I}\Omega$ is the Hessian matrix of $\Omega$. Since the order parameters minimize $\Omega$, the matrix $\partial^2_{\Delta^I}\Omega$ is positive semidefinite (it is actually positive definite since we fixed the overall phase of the order parameters). Thus $({\rm d}_i\Delta^I)^{\rm T}\partial^2_{\Delta^I}\Omega({\rm d}_i\Delta^I)\geq 0$, and Eq.~\eqref{eq.inequality} follows from Eq.~\eqref{eq.sfw_trs}.   
This implies that using $[D_s]_{ij}=(e^2/V\hbar^2)\partial^2\Omega/\partial q_i\partial q_j\big|_{\vec{q}=\vec{0}}$ overestimates the diagonal components of the superfluid weight, and can even predict a nonzero superfluid weight when it is really zero~\cite{Huhtinen2022}. 

\subsection{Superfluid weight from linear response: the conventional and geometric contributions} \label{sec.lin_resp}
	
	As explained in Sec.~\ref{sec.sf_weight}, the superfluid weight can be defined both in terms of the free energy and, equivalently, the current induced by a vector potential $\vec{A}$. That is, the superfluid weight can also be defined as the following static transport coefficient~\cite{Scalapino1993}:
	\begin{equation}
		\label{eq:superfluid_weight}
		[D_{{\rm s}}]_{ij}  = -\lim_{\vec{q}_\perp \to 0} \chi_{ij}(q_{\parallel}=0,\vec{q}_{\perp},\omega = 0)\,,
	\end{equation}
	where $\chi_{ij}(\vec{q},\omega)$ is the current-current response function, i.e. the current induced by a vector potential $\vec{A}(\vec{q},\omega)$ is $j_{i} = \sum_{j}\chi_{ij}A_{j}$. In the above equation, the wavevector $\vec{q} = q_\parallel \hat{j} + \vec{q}_\perp$ is decomposed into the colinear (${\rm q}_{\parallel}$) and perpendicular ($\vec{q}_\perp$) components with respect to the $j = x,y,z$ axis (note that $j$ is the second index appearing in $[D_{s}]_{ij}$ and $\chi_{ij}$). The vector potential enters the non-interacting Hamiltonian $H_0$ (see Eq.~\eqref{eq.H0}) via the Peierls substitution~\cite{Scalapino1993}. In the exact Hubbard model, $H_{\rm int}$ does not contain $\vec{A}$, but in the mean-field approximation Eq.~\eqref{eq.Hint_mf}, the vector potential enters $H_{\rm int}$ through the order parameters, which need to be solved for each $\vec{A}$. 
	
	The calculation of the superfluid weight in mean-field theory via linear response can be found in Appendix D of Ref.\cite{Huhtinen2022}. Here we just give the result:
\begin{equation}
\label{eq.linrespresult}
\begin{split}
  [D_s]_{ij} = \frac{e^2}{V\hbar^2}&
  \sum_{\vec{k},ab}\frac{n_F(E_a)-n_F(E_b)}{E_b-E_a}\big[ 
    \bra{\psi_a}\partial_{k_i}\widetilde{H}_{\vec{k}} \ket{\psi_b}
    \bra{\psi_b} \partial_{k_j}\widetilde{H}_{\vec{k}} \ket{\psi_a}  \\
    -&\bra{\psi_a}(\partial_{k_i}\widetilde{H}_{\vec{k}}\gamma^z+\delta_{i}\Delta)
    \ket{\psi_b} \bra{\psi_b}
    (\partial_{k_j}\widetilde{H}_{\vec{k}}\gamma^z
    +\delta_{j}\Delta)  
    \ket{\psi_a} 
    \big] - \frac{1}{V_c}C_{ij}, 
    \end{split}
    \end{equation}  
where
\begin{equation}
\begin{split}
  \partial_{k_i}\widetilde{H_{\vec{k}}} &= \begin{pmatrix}
    \frac{\partial H_{\vec{k}'}^{\up}}{\partial
      k_{i}'}\bigg|_{\vec{k'}=\vec{k}} & 0 \\
    0 & \frac{\partial (H_{\vec{k}'}^{\dn})^*}{\partial
      k_{i}'}\bigg|_{\vec{k'}=-\vec{k}}
  \end{pmatrix},  \\
  \delta_{i}\Delta &= \begin{pmatrix}
    0 & \frac{{\rm d}\vec{\Delta}}{{\rm d}q_{i}}\bigg|_{\vec{q}=\vec{0}} \\
    \frac{{\rm d}\vec{\Delta}^{\dag}}{{\rm d}q_{i}}\bigg|_{\vec{q}=\vec{0}} & 0
  \end{pmatrix},  \\
  C_{ij} &= \frac{e^2}{U\hbar^2}\sum_{\alpha} \frac{{\rm d}\Delta_{\alpha}}{{\rm d}q_i}\frac{{\rm d}\Delta_{\alpha}^*}{{\rm d}q_j}\bigg|_{\vec{q}=\vec{0}} + {\rm H.c.}
  \end{split}
\end{equation}
We define $\gamma^z=\sigma_z\otimes \mathbb{1}_{N_b}$, with $\sigma_i$ the Pauli matrices. The Fermi-Dirac distribution at energy $E$ is $n_{F}(E) = 1/(e^{\beta E}+1)$, and $E_a$ and $\ket{\psi_a}$ are the eigenvalues and eigenvectors of $H_{\rm BdG}$. When $E_a=E_b$, the prefactor in~\eqref{eq.linrespresult} is $-\partial n_F(E_a)/\partial E_a$. The derivatives ${\rm d}\Delta_{\alpha}/{\rm d}q_i\big|_{\vec{q}=\vec{0}}$, $\vec{q}$ are the same as defined in Sec.~\ref{sec.mf_sfw} (where $\vec{q}$ enters the Hamiltonian through the transformation $\Delta_{i\alpha}\to \Delta_{i\alpha}e^{2i\vec{q}\cdot \vec{r}_{i\alpha}}$), and can be computed for example by solving the system of equations Eq.~\eqref{eq.sys_del_mu}. Here, $\mu$ is assumed constant, and Eq.~\eqref{eq.linrespresult} is equivalent to Eq.~\eqref{eq.sfw_full} whenever ${\rm d}\mu/{\rm d}q_i\big|_{\vec{q}=\vec{0}}$. If we had ignored the derivatives of the order parameters, we would have obtained a result equivalent to $(e^2/V\hbar^2)\partial^2\Omega/\partial q_i\partial q_j\big|_{\vec{q}=\vec{0}}$. Note that in the literature before Ref.~\cite{Huhtinen2022}, the derivatives of the order parameters are typically ignored. This is in general not correct; although in many cases where the system is highly symmetric they are actually zero, neglecting them may in some other cases produce qualitatively incorrect results such as prediction of superfluidity in a disconnected system. Therefore, one should, in general, apply the full formulas presented in these lecture notes and in Ref.~\cite{Huhtinen2022}. Ref.~\cite{Peotta2022} shows how the terms related to the derivatives of the order parameters can be understood as a  random-phase-approximation (RPA) type approach to the calculation of the superfluid weight.

To understand why the superfluid weight can be nonzero even in a flat band, it is instructive to split $D_s$ into so-called conventional and geometric contributions. We will focus here on systems with TRS. To express $D_s$ in terms of the Bloch functions, we write $\ket{m_{\sigma,\vec{k}}}$: $\ket{\psi_a}=\sum_{m=1}^{n}
(w_{+,am} \ket{+}\otimes\ket{m_{\up,\vec{k}}}+w_{-,am}\ket{-}\otimes\ket{m^*_{\dn,-\vec{k}}})$, where
$\ket{m_{\up,\vec{k}}}$ is the eigenvector of $H_{\vec{k}}^{\up}$ with
eigenvalue $\epsilon_{\up,m,\vec{k}}$, $\ket{m^*_{\dn,-\vec{k}}}$ is the
eigenvector of $(H_{-\vec{k}}^{\dn})^*$ with eigenvalue
$\epsilon_{\dn,m,-\vec{k}}$, and $\ket{\pm}$ are the eigenvectors of $\sigma_z$ with eigenvalues $\pm 1$. Then, we split the superfluid weight in Eq.~\eqref{eq.linrespresult} as $D_s=D_{s,{\rm conv}}+D_{s,{\rm geom}}$, where the conventional part is
\begin{equation}
\begin{split}
  &[D_{s,{\rm conv}}]_{ij} = \frac{e^2}{V\hbar^2}\sum_{\vec{k}}\sum_{mn} C_{nn}^{mm}
  [j_{i}^{\up}(\vec{k})]_{mm} [j_{j}^{\dn}(-\vec{k})]_{nn},  \\
  &C_{pq}^{mn} = 4 \sum_{ab} \frac{n_F(E_a)-n_F(E_b)}{E_b-E_a}
  w_{+,am}^*w_{+,bn} w_{-,bp}^* w_{-,aq},  \\
  &[j_{i}^{\sigma}(\vec{k})]_{mn} = \bra{m_{\sigma,\vec{k}}}
  \partial_{k_{i}} H_{\vec{k}}^{\sigma} \ket{n_{\sigma,\vec{k}}}  \\
  &\phantom{[j_{i}^{\sigma}(\vec{k})]_{mn}}= \delta_{mn}\partial_{k_{i}}\epsilon_{\sigma,m,\vec{k}} + (\epsilon_{\sigma,m,\vec{k}}-\epsilon_{\sigma,n,\vec{k}}) \braket{\partial_{k_{i}}m_{\sigma,\vec{k}}}{n_{\sigma,\vec{k}}}.
\end{split}
\end{equation}
This contribution contains only single-band components of the current operators, and is the only contribution to the superfluid weight in single-band models. Since the dispersion of a flat band is constant, $D_{s,{\rm conv}}$ is clearly zero on flat bands.

The geometric contribution $D_{s,{\rm geom}}$ can be split into $D_{s,{\rm geom}}=D_{s,{\rm geom}}^1+D_{s,{\rm geom}}^2+D_{s,{\rm geom}}^3$, with
\begin{align}
    [D_{s,{\rm geom}}^1]_{ij} &= 
    \frac{e^2}{V\hbar^2}\sum_{\vec{k}}\sum_{\substack{m\neq n \\ p\neq q}} C_{pq}^{mm}
  [j_{i}^{\up}(\vec{k})]_{mn} [j_{j}^{\dn}(-\vec{k})]_{qp}, \\
  \begin{split}
  [D_{s,{\rm geom}}^2]_{ij} &= 
  \frac{e^2}{V\hbar^2}\sum_{\vec{k}}\sum_{\substack{m\\p\neq q}} C_{pq}^{mm}
  [j_{i}^{\up}(\vec{k})]_{mm} [j_{j}^{\dn}(-\vec{k})]_{qp} \\&+ \frac{e^2}{V\hbar^2}\sum_{\vec{k}}\sum_{\substack{m\neq n\\ p}} C_{pp}^{mn}
  [j_{i}^{\up}(\vec{k})]_{mn} [j_{j}^{\dn}(-\vec{k})]_{pp},
  \end{split}\\
  \begin{split}
  [D_{s,{\rm geom}}^3]_{ij} &= -\frac{e^2}{V\hbar^2} \sum_{\vec{k},ab}
  \frac{n_F(E_a)-n_F(E_b)}{E_b-E_a} 
  \big( \bra{\psi_a}\delta_{i}\Delta^I \ket{\psi_b}\bra{\psi_b}\delta_{j}\Delta^I\ket{\psi_a}
  \\
  & 
  + \bra{\psi_a}\delta_{i}\Delta^I\ket{\psi_b}
  \bra{\psi_b}\partial_{k_j} \widetilde{H}_{\vec{k}}\gamma^z \ket{\psi_a}
  + \bra{\psi_a}\partial_{k_{i}}\widetilde{H}_{\vec{k}}\gamma^z\ket{\psi_b}\bra{\psi_b} \delta_{j}\Delta^I\ket{\psi_a}\big)
  \\&-\frac{2e^2}{UV_c\hbar^2}\sum_{\alpha} \frac{{\rm d}\Delta_{\alpha}^I}{{\rm d}q_{i}} \frac{{\rm d}\Delta_{\alpha}^I}{{\rm d}q_{j}}\bigg|_{\vec{q}=\vec{0}},
  \end{split}
\end{align}
where 
\begin{equation}
\delta_{i}\Delta^I = \begin{pmatrix}
0 & i\dfrac{{\rm d}({\rm Im}(\vec{\Delta}))}{{\rm d}q_{i}}\bigg|_{\vec{q}=\vec{0}} \\
-i\dfrac{{\rm d}({\rm Im}(\vec{\Delta}))}{{\rm d}q_{i}}\bigg|_{\vec{q}=\vec{0}} & 0
\end{pmatrix}.
\end{equation}
The contribution $D_{s,{\rm geom}}^3$ contains all the terms involving derivatives of the imaginary parts of the order parameters in Eq.~\eqref{eq.linrespresult}. As the derivatives of the real parts of the order parameters are zero in the presence of TRS, $D_{s, {\rm conv}}+D_{s,{\rm geom}}=D_s$ with the definitions above. The contribution $D_{s,{\rm geom}}^3$ is included in the geometric contribution because it can only be nonzero in a multiband system: for a single band, due to the freedom in the phase of the order parameter, ${\rm d}\Delta^I/{\rm d}q_i\big|_{\vec{q}=\vec{0}}$ can always be set to zero by simply keeping it real and positive at all $\vec{q}$. The geometric contribution is only present in multiband systems, and can be nonzero even on a flat band. We will see in the next section that it is related to the minimal quantum metric in isolated flat bands with uniform pairing~\cite{Huhtinen2022}. 

The above definitions are valid in a system with TRS, where the derivatives of the order parameters that enter the superfluid weight are purely imaginary. In a system without TRS, Eq.~\eqref{eq.linrespresult} also contains terms involving the derivatives of the real parts of the order parameters, which can be nonzero even in single-band models in contrast to ${\rm d}\Delta^I/{\rm d}q_i\big|_{\vec{q}=\vec{0}}$. In that case, the definition of $D_{s,{\rm conv}}$ should be adapted so that it also contains those terms involving ${\rm d}\Delta_{\alpha}^R/{\rm d}q_i\big|_{\vec{q}=\vec{0}}$ which can be nonzero in a single-band model. Note that $D_{s,{\rm geom}}$ will also have additional contributions containing ${\rm d}\Delta_{\alpha}^R/{\rm d}q_i\big|_{\vec{q}=\vec{0}}$, for example those which additionally involve ${\rm d}\Delta_{\alpha}^I/{\rm d}q_i\big|_{\vec{q}=\vec{0}}$.

\section{Quantum metric and isolated flat bands} \label{sec.sfw_fb}

    In this Section, we show how quantum geometry is related to the superfluid weight in flat band systems. We focus on an isolated flat band with TRS, which means that there is a large band gap $E_{\rm gap}$ between the partially filled flat band and all other bands
	\begin{equation}
		\label{eq:isol_band_lim}
		|\varepsilon_{m\vc{k}}-\varepsilon_{\bar{n}\vc{k}}| \gtrsim E_{\rm gap} \gg \Delta \quad \text{for} \quad m\neq \bar{n}\,.
	\end{equation}
	Furthermore, we assume that the uniform pairing condition is fulfilled, implying that $\Delta_{\alpha}$ is equal at all orbitals where the flat band states reside, and zero otherwise.

The superfluid weight in such systems is given by
\begin{align}
  [D_{s}]_{ij}  &= \frac{4e^2}{(2\pi)^{D-1}\hbar^2 N_{\rm orb}}|U|f(1-f)\mathcal{M}^{\rm min}_{ij}, \label{eq.Ds_qm}\\
  \mathcal{M}^{\rm min}_{ij} &= \frac{1}{2\pi} \int {\rm d}^D\vec{k} \, {\rm
    Re }(\mathcal{B}_{ij}(\vec{k}))_{\rm min},
\label{eq:qmetricconnetion}
\end{align}
Here, $f$ is the filling fraction of the isolated flat band, $N_{\rm orb}$ is the number of orbitals where the flat band states have a nonzero amplitude and $\mathcal{B}_{ij}(\vc{k})$ is the quantum geometric tensor introduced in Eq.~\eqref{abelianQGT}.
Here, $P(\vec{k})=\ket{\bar{n}_{\vec{k}}}\bra{\bar{n}_{\vec{k}}}$ is the projector into the Bloch state of the isolated band $\bar{n}$. The label "min" refers to the integrated quantum metric whose trace is minimal with respect to possible choices of orbital positions (i.e., the positions of lattice sites in the unit cell, assuming the same hoppings). Note that the superfluid weight is independent of orbital positions, and so is the minimal quantum metric. Originally, the connection between flat band superconductivity, quantum metric and Berry curvature was made in~\cite{Peotta2015,Liang2017}, where the result is of the same form as Eq.~(\ref{eq:qmetricconnetion}), but with just the quantum metric, not the minimal one. That has the problem that the quantum metric, as well as Berry curvature, actually depend on the orbital positions. The discrepancy was resolved in Ref.~\cite{Huhtinen2022} by showing the dependence on the minimal quantum metric; for details of that calculation, see Appendix A. 

It may seem that minimizing the quantum metric with respect to all orbital positions might be a tedious numerical task, at least in complicated lattices. Fortunately, in Ref.~\cite{Huhtinen2022} it was also shown that the orbitals should be placed at high-symmetry positions for the quantum metric to be minimized. If unique high-symmetry positions exist for the model, the quantum metric will thus automatically be minimized when they are chosen as the orbital positions.

The above results are valid in systems with TRS. For discussion of the systems where TRS is broken, see Appendix A.

     The natural interpretation of 
     Eq.~\eqref{eq.Ds_qm} is that the effective mass appearing in the standard expression for the superfluid weight $D_{\rm s} = \frac{e^2 n}{m_{\rm eff}}$ is not the one obtained from the single-particle band structure, but rather the effective mass of the two-body bound state~\eqref{eq:eff_mass_flat} and~\eqref{eq:eff_mass_flat_qm} as discussed in Sec.~\ref{sec:two_body}~\cite{Torma2018,Iskin2021twobody}.
    In Eqs.~\eqref{eq:eff_mass_flat} and~\eqref{eq:eff_mass_flat_qm}, the quantum metric instead of the minimal one appears but, as said above, if one chooses the orbitals to be at the high-symmetry positions, the result will be the same. Moreover, in Ref.~\cite{Torma2018}, Eq.~\eqref{eq:eff_mass_flat_qm} was obtained after some approximations and the non-approximate result was actually orbital-independent. In Ref.~\cite{Huhtinen2022}, it is shown that the two-body effective mass is in fact related to the minimal quantum metric, just like the superfluid weight. 
	 
    For non-flat band superconductors 
    one can often safely ignore the geometric contribution since it is an effect of order $\Delta$, while the conventional one is proportional to the Fermi energy $E_{\rm F}$, a much larger energy scale, which is roughly set by the bandwidth of the partially filled band. There are few cases in which the ratio $\Delta/E_{\rm F}$ is large and therefore the geometric contribution becomes important, among which is the recently discovered superconducting state in magic-angle TBG~\cite{MacDonald2019,Andrei2020,Balents2020,Kennes2021,Andrei2021}. The need to take into account the off-diagonal (inter-band) matrix elements of the current operator was put forward for the first time in the context of the exciton superfluid phase occurring in quantum Hall bilayers~\cite{Moon1995}, which may be regarded as the first flat band superfluid to be experimentally realized.
    On the theoretical side, Kopnin computed the superfluid weight of the flat band of surface states of rhombohedral graphene showing that it is nonzero~\cite{Kopnin2011a}. 
    This result was important to  validate the concept of flat band superfluidity and its promise to increase the temperature of the superconducting transition. Indeed, the superfluid weight is directly related to the critical temperature in two-dimensional superfluids (see Sec.~\ref{sec:BKT_temperature}). However, it was unclear what specific properties of the flat band wave functions determine the superfluid weight and whether this result can be extended to other lattices. The connection between superfluidity and quantum geometry~\cite{Peotta2015,Liang2017}  embodied by~\eqref{eq.Ds_qm}  is the general answer to this question.

	\section{The physical interpretation and the Wannier function overlaps}
	\label{sec:Wannier_overlap}

	In Sections~\ref{sec:two_body} and~\ref{sec.sfw_fb} we showed that the quantum metric determines the superfluid weight of a flat band superconductor. Thus it is robust against  thermal fluctuations, and one also expects the critical current to scale with the superfluid weight. 
	The quantum metric is a rather abstract concept, thus it is natural to ask what it means for a flat band to have a large quantum metric and why this leads to a large superfluid weight. An answer to this question can be formulated using the idea of Wannier functions.

	A Wannier function  is defined as the Fourier transform of the periodic Bloch function 
    \begin{equation*}
     w_n(\vc{i},\alpha) =\frac{V_{\rm c}}{(2\pi)^2}\int {\rm d}^D\vc{k}\,e^{i\vc{k}\cdot\vc{r}_{\vc{i}\alpha}}n_{\vc{k}}(\alpha)   
    \end{equation*}
    and the set $\{w_n(\vc{i}-\vc{j},\alpha)\}_{\vc{j}}$ obtained by shifting this Wannier function across the lattice is an orthonormal basis of the subspace corresponding to the $n$-th band, which is assumed to be isolated. They satisfy  orthonormality 
    \begin{equation*}
     \sum_{\vc{i}\alpha} w_n^*(\vc{i}-\vc{j},\alpha)w_n(\vc{i}-\vc{l},\alpha) = \delta_{\vc{j},\vc{l}},   
    \end{equation*}
    and completeness 
    \begin{equation*}
    \sum_{\vc{l}}w_n(\vc{i}-\vc{l},\alpha)w_n^*(\vc{j}-\vc{l},\beta) = P_{\alpha,\beta}(\vc{i}-\vc{j}),    
    \end{equation*}
    where $P_{\alpha,\beta}(\vc{i}-\vc{j})$ is the real space representation of the projector on the $n$-th band subspace. There is no canonical choice for the Wannier functions since one can multiply the Bloch functions by an arbitrary phase, that is $\ket{n_{\vc{k}}} \to e^{i\theta(\vc{k})}\ket{n_{\vc{k}}}$, a procedure which leads to new Wannier functions which can be different from the original ones. Finding suitable criteria to restrict this gauge freedom in the choice of Wannier functions has been the subject of considerable work. A common prescription is the one based on the Marzari-Vanderbilt localization functional, which is a quantity measuring how much the Wannier functions are spread or delocalized in real space~\cite{Marzari1997}. The minimization of the Marzari-Vanderbilt functional leads to the maximally localized Wannier functions, which have found extensive applications in electronic structure theory~\cite{Marzari2012}.
	
	The Marzari-Vanderbilt functional is the sum of two positive parts, one which depends on the gauge choice for the Wannier functions, while the other is gauge-invariant, i.e. is unchanged upon multiplication of the Bloch functions by an arbitrary phase factor. The gauge-invariant part is the trace of the quantum metric integrated over the BZ, which sets a lower bound on how much the Wannier functions can be localized. 
	Therefore, one way to think about the quantum metric is as quantity measuring the overlap/spread of the localized wave function of the flat band.

\begin{figure}
\begin{minipage}[s]{\textwidth}
\justifying
\includegraphics[width=0.8\columnwidth]{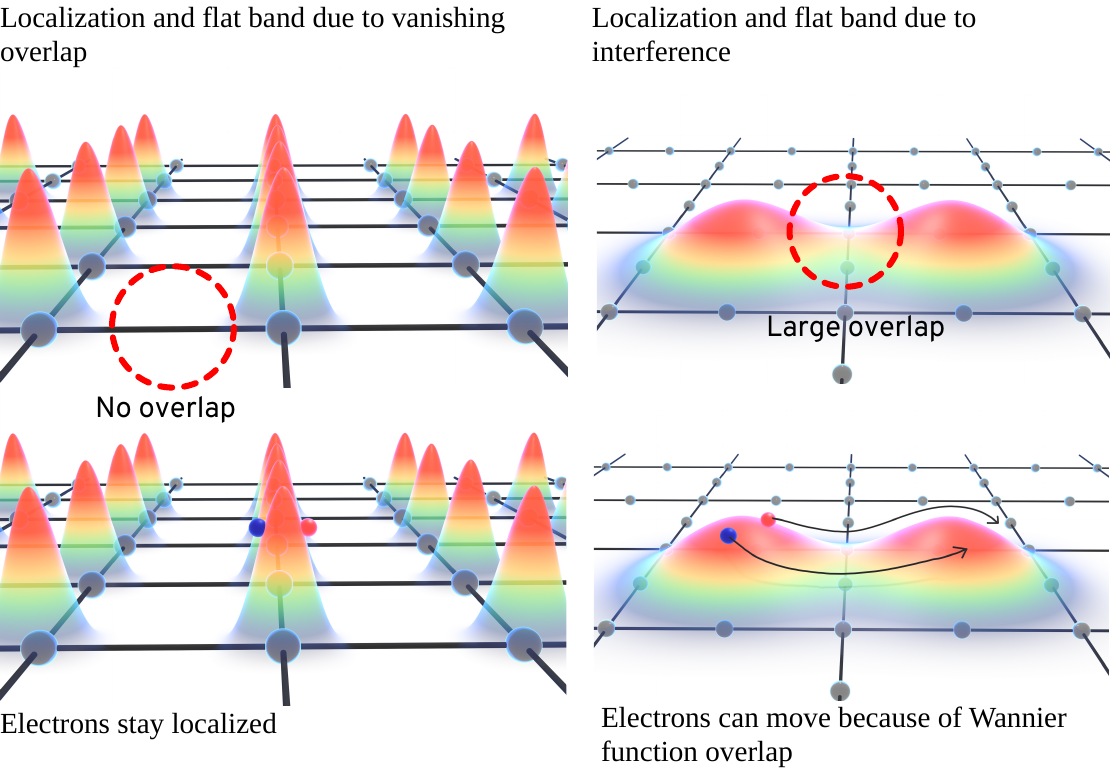}
\caption{\label{fig:figure1} There are two ways for non-interacting particles can be immobile: their Wannier functions can be exponentially localized at the lattice sites, or they can be localized due to destructive interference. In the former case, interacting particles will remain localized too, while in the latter, the interference is distorted by the interactions and the particle movement is facilitated by the finite overlap of the Wannier functions. Wannier function overlaps are determined by the quantum geometry of the band and exponential localization of the Wannier functions is impossible in topologically non-trivial systems. Picture credits Antti Paraoanu.}
\end{minipage}
\end{figure}

In a topologically nontrivial band there is an obstruction to the full localization of Wannier functions. The first, most  important, and simplest topological invariant is the Chern number~\cite{Chiu2016} $\mathcal{C}$ defined in Eq.~\ref{eq:Chernnumber}.   The positive semidefiniteness of the QGT implies~\cite{Peotta2015,Huhtinen2022}
	\begin{equation}
		\label{eq:ineq}
		\mathrm{det}\,\mathcal{M}^{\rm min} \geq \mathcal{C}^2\,,\,\,\text{with}\,\,\mathcal{M}^{\rm min}_{ij} =\frac{1}{2\pi}\int {\rm d}^2\vc{k}\,\mathrm{Re}\,\mathcal{B}_{ij}(\vc{k})_{\rm min}\,. 
	\end{equation}
	This, together with the fact that the matrix $\mathcal{M}^{\rm min}$ appears also in Eq.~\eqref{eq.Ds_qm}, gives a lower bound on the superfluid weight of a topologically nontrivial flat band (time-reversal symmetry is assumed here, therefore $\mathcal{C}$ is the spin Chern number). This is consistent with the fact that it is not possible to construct exponentially localized wave functions in a band with nonzero Chern number~\cite{Brouder2007,Panati2007}, the decay can be at best an algebraic one of the form $1/r^2$~\cite{Marcelli2019arxiv,*Marcelli2019journal,Monaco2016,Monaco2018}. Slowly decaying Wannier functions necessarily have a large overlap, which is expected to lead to a robust superfluid state. A precise connection between Wannier functions overlap and superfluid order has been worked out in Ref.~\cite{Tovmasyan2016}, where it has been shown that  the interaction Hamiltonian projected on the flat band can be mapped onto an effective spin Hamiltonian, whose exchange couplings are precisely the Wannier function overlaps. Thus, the larger the Wannier function overlap the more robust is the superfluid order, which  is represented as ferromagnetic order according to the mapping into the effective spin Hamiltonian. The role of the Wannier function overlap in flat band superfluidity and superconductivity is illustrated in Fig.~\ref{fig:figure1}.
	
	Inequalities between the quantum metric integrated over the BZ, which is the invariant part of the Marzari-Vanderbilt localization functional, and topological invariants are known in the case of the winding number in $D=1$~\cite{Tovmasyan2016} and the Euler class in $D = 2$~\cite{Xie2020}. This latter invariant has recently found application  in magic-angle twisted bilayer graphene (MA-TBG)~\cite{Torma2022} and provides a lower bound for the superfluid weight of the superconducting phase occurring in this material. One can reasonably expect similar inequalities to exist for any topological invariant, but they have not been obtained yet.
	
	Wannier functions have many important applications which generally rely on their localization properties~\cite{Marcelli2020}. For instance, they are used to construct effective tight-binding Hamiltonians that capture the relevant degrees of freedom of a material and thus reduce the single-particle Schr\"odinger equation in the continuum to a more manageable discrete one~\cite{Marzari2012}. The Peierls substitution, in which the vector potential is added as a phase to the hoppings,
	is generally used to model the interaction with the electromagnetic field; however this is an approximation which is not easy to justify in the case of a real material~\cite{Alexandrov1991,Boykin2001,Boykin2001a,Graf1995,Kohn1959,Luttinger1951}. The first rigorous results on the existence of exponentially localized Wannier functions were driven by the need to justify the Peierls substitution~\cite{Kohn1958,Kohn1959,Kohn1959a}. Only relatively recently has it been shown that the Chern number is the only obstruction to the existence of exponentially localized Wannier functions in $D\leq 3$ and in the absence of other unitary or anti-unitary symmetries~\cite{Panati2007,Brouder2007}. From this result it follows that in the effective tight-binding model one should include enough bands so that the total Chern number (the sum of the Chern numbers of each band) is zero, otherwise the Peierls substitution is not justified and the effective model is of little use. Therefore it becomes clear that to describe the superfluid properties of a topologically non-trivial ($\mathcal{C}\neq 0$) flat, or quasi-flat, band a multiband approach, such as the one in Sec.~\ref{sec.conv_geom}, is unavoidable. This is also the case of a topologically trivial flat band with a large quantum metric since a single-band effective model would give a zero result for the superfluid weight in contrast to Eq.~\eqref{eq.Ds_qm}.
	
	\section{The BKT temperature}
	\label{sec:BKT_temperature}
	
	
	One of the main reasons behind the interest in materials with quasi-flat bands in the band structure is that the critical temperature of the superconducting transition is strongly enhanced due to the high density of states~\cite{Kopnin2011,Heikkila2011,Khodel1990,Khodel1994}. Indeed, according to the BCS theory the critical temperature scales as $T_{\rm c} \sim \exp({-1/|U|\rho(\varepsilon_{\rm F})})$ in the case where the bandwidth is much larger than the interaction energy scale~\cite{Schrieffer1964}.  Instead, in a perfectly flat band the critical temperature is proportional to the interaction strength $T_{\rm c} \sim |U|$. This is essentially the same result as for the binding energy of the two-body bound state discussed in Sec.~\ref{sec:two_body}, meaning that, within simple BCS theory,  the transition is driven by the breaking of Cooper pairs due to thermal excitations. 
	
	While this picture is accurate in the weak-coupling limit, it leads to an unphysical unbounded growth of the critical temperature in the strong-coupling limit.~\cite{Liang2017} The order parameter is a static quantity at the mean-field level, but in reality it is affected by (thermal) fluctuations governed by the superfluid weight. The latter can be computed by introducing a modulation of the order parameter (this approach is equivalent to the one based on linear response theory of Sec.~\ref{sec.lin_resp}), therefore it can also be used to estimate the critical temperature. In $D=2$ the relation between critical temperature and superfluid weight is given by the universal relation $T_{\rm BKT} = \frac{\pi \hbar^2}{8e^2}\sqrt{{\rm det}[D_{\rm s}(T_{\rm BKT})]}$~\cite{Nelson1977}, where $T_{\rm BKT}$ is the Berezinskii–Kosterlitz–Thouless (BKT) temperature~\cite{Berezinskii1971,Berezinskii1972,Kosterlitz1973}. When the superfluid weight is diagonal with $[D_{\rm s}]_{xx}=[D_{\rm s}]_{yy}$, which is for example the case in rotationally symmetric systems, this reduces to the simpler form $T_{\rm BKT} = \frac{\pi \hbar^2}{8e^2}[D_{\rm s}(T_{\rm BKT})]_{xx}$
	The critical temperature  of a two-dimensional superfluid  is thus proportional to the superfluid weight at the critical temperature $D_{\rm s}(T_{\rm BKT})$.
	As in the case of the BCS critical temperature, the BKT temperature is proportional to the interaction strength $T_{\rm BKT}\sim |U|$ in the isolated flat band limit. However, it reaches a maximum for large enough $U$ and in the strongly coupling limit it decreases as $|U|^{-1}$. Indeed, in this limit the Cooper pairs are tightly bound and their effective hopping amplitude, proportional to the inverse of the pair effective mass, is $J \sim t^2/|U|$, with $t$ the single particle hopping, as can be shown with a strong coupling expansion~\cite{Liang2017,Ho2009,Nozieres1985}. 
	
    By comparison with other quantum many-body methods, such as dynamical mean-field theory~\cite{Julku2016,Liang2017}, quantum Monte Carlo~\cite{Hofmann2020,Peri2021,Herzog-Arbeitman2021}, density matrix renormalization group~\cite{Chan2021,Tovmasyan2018,Mondaini2018}, and exact analytical calculations~\cite{Herzog-Arbeitman2022}, it has been verified that the predictions of mean-field theory are  generally accurate, in particular the linear dependence of the superfluid weight and the critical temperature with the interaction strength is a robust result for $|U|\lesssim E_{\rm gap}$. Intriguingly, the fact that the superconducting gap is an analytic function of the coupling constant in the flat band limit implies that superconductivity can be studied using perturbation theory. This is not the case for a dispersive band due to the non-analytic dependence $\sim e^{-1/x}$ of the superconductive gap. The perturbative approach for flat band superconductors has been developed to some extent in Ref.~\cite{Tovmasyan2016}.

\section{Quantum geometry and Bose-Einstein condensation}

Using the Bogoliubov theory of weakly interacting Bose-Einstein condensation (BEC), it is easy to show that the quantum depletion (the fraction of excitations out of the condensate created by the interactions) diverges in a single flat band. Again, however, quantum geometry helps stabilize the BEC. In Ref.~\cite{Julku2021,Julku2021b} it was shown that
the speed of sound in a flat band (multiband) BEC is proportional to the quantum metric, and the quantum depletion to the quantum distance between Bloch states in the flat band. The superfluid weight was shown to be dominated by quantum fluctuations. The need to ensure orbital-position-independent results, as shown in the fermionic superconductivity case in~\cite{Huhtinen2022}, is considered in the BEC case in Ref.~\cite{Julku2023}.

\section{Conclusions} 

Quantum geometric effects on superfluidity and superconductivity have the most remarkable impact in (quasi-)flat bands. 
Flat and quasi-flat bands can be experimentally realized in several solid state, ultracold gas and photonic systems~\cite{Leykam2018}. Moir\'e materials have recently emerged as a highly promising platform for flat band studies, highlighted by the observation of superconductivity in twisted bilayer graphene~\cite{MacDonald2019,Andrei2020,Balents2020,Kennes2021,Andrei2021}; also in ultracold gases~\cite{Bloch2008,Giorgini2008,Torma2015,Lewenstein2012} work on moir\'e systems has started. For a review on the potential role of quantum geometric superconductivity in graphene and moir\'e optical lattices see Ref.~\cite{Torma2022}. In twisted bilayer graphene systems, first experiments studying the role of quantum geometry in superconductivity have been recently reported~\cite{Tian2023}.  
To unambiguously observe quantum geometric flat band superconductivity, one would need to verify the linear dependence on the interaction of the Cooper pairing temperature, the pairing gap, superfluid weight, and the BKT superfluid transition temperature, for interactions smaller than or similar to the hopping energies. Another clear signature is the dependence of these quantities on the filling of the band: quantum geometric superconductivity maximizes at half filling (see Eq.~\eqref{eq.Ds_qm}), in contrast to the single band case where higher fillings or density of states singularities mean stronger superfluidity. Additionally, the Josephson effect could show effects unique to flat band superfluidity~\cite{Pyykkonen2021}. Apart from the superconducting state, the normal state above it in temperature is expected to be interesting in flat bands: for instance preformed pairs, pseudogaps, and insulator behavior characteristic for a flat band have been predicted~\cite{Tovmasyan2018,jiang_charge_2019,WangLevin2020,Peri2021,Huhtinen2021}.

The quantum metric quantifies distances between the eigenstates of a system, thus it naturally influences many observables of interacting systems. Light-matter interactions~\cite{Topp2021,Chaudhary2021} and exciton condensates~\cite{hu2020excitoncondensate} in \MR{}~materials have been predicted to be influenced by the band quantum geometry. As mentioned above, quantum geometry was predicted to be important for the stability of BEC in flat bands~\cite{Julku2021,Julku2021b}, which is relevant for bosonic condensates in ultracold gas and polariton systems, and also 2D \MR{} materials when they are in the bosonic side of the BCS-BEC crossover~\cite{Park2021}.

Probably the biggest promise of quantum geometric superconductivity is related to the potential of higher critical temperature. Both the existence~\cite{Hazra2019,Verma2021} and absence~\cite{Hofmann2021} of an upper bound for the critical temperature has been predicted. Based on the fundamental lower bound of superfluidity Eq.~\eqref{eq:ineq}, higher critical temperatures could be expected for a large quantum metric or Chern number. While these results assume the isolated flat band limit, having other bands touching the flat band can lead to an even higher critical temperature than in the case of an isolated flat band~\cite{Julku2016,Huhtinen2022}. This is promising as many real materials have flat bands with other bands nearby in energy. Discovering simple analytical, approximate results for the critical temperature and superfluid weight also in the non-isolated flat band case would provide an invaluable guideline in the search of superconductivity at higher temperatures, eventually up to room temperature.

\vspace{2\baselineskip}

\acknowledgments
We acknowledge support by the Academy of Finland under project numbers 303351, 327293, 349313, 330384 and 336369. K-E.H. acknowledges support from the Magnus Ehrnrooth Foundation. 

\newpage
\noindent\textsc{Appendix A.}

\vspace{0.7cm}
In this appendix, we explain why the superfluid weight is related to the minimal quantum metric, i.e. the one with the smallest possible trace.
If the terms relating to the derivatives of the order parameters are ignored, the superfluid weight in isolated flat bands with uniform pairing is given by~\cite{Peotta2015,Liang2017}
\begin{equation}
\label{eq.Ds_qm_bis}
\begin{split}
  [D_{s}]_{ij}  &= \frac{4e^2}{(2\pi)^{D-1}\hbar^2 N_{\rm orb}}|U|f(1-f)\mathcal{M}_{ij}, \\
  \mathcal{M}_{ij} &= \frac{1}{2\pi} \int_{\rm B.Z.} {\rm d}^D\vec{k} \, {\rm
    Re }(\mathcal{B}_{ij}(\vec{k})),
\end{split}
\end{equation}
Here, $f$ is the filling fraction of the isolated flat band, $N_{\rm orb}$ is the number of orbitals where the flat band states have a nonzero amplitude, and the integral is over the first Brillouin zone. As stated in the main text, $\mathcal{B}_{ij}$ is the quantum geometric tensor
\begin{equation}
\mathcal{B}_{ij}(\vec{k}) = 2 \text{Tr }P(\vec{k}) \partial_{k_i} P(\vec{k})\partial_{k_j} P(\vec{k}).
\end{equation}
Here, $P(\vec{k})=\ket{\bar{n}_{\vec{k}}}\bra{\bar{n}_{\vec{k}}}$ is the projector into the Bloch state of the isolated band $\bar{n}$.

It is important to stress here that this equation is only valid when the derivatives of the order parameters ${\rm d}\Delta_{\alpha}/{\rm d}q_i\big|_{\vec{q}=\vec{0}}$ appearing in Eq.~\eqref{eq.sfw_trs} are zero. As explained in the main text, this is not generally the case in multiband models, and it is thus not immediately clear when Equation~\eqref{eq.Ds_qm} can be used. An important parameter to understand the conditions under which Eq.~\eqref{eq.Ds_qm} is valid is the choice of intra-unit cell positions $\{\vec{\delta}_{\alpha}\}$. This choice affects the quantum metric $\mathcal{M}$ through the convention of Fourier transformation Eq.~\eqref{eq.fourier}, but does not change the free energy, and therefore has no effect on $D_s$: clearly, when relating $\mathcal{M}$ to $D_s$, we need some condition for $\{\vec{\delta}_{\alpha}\}$. We will now show that (1) intra-unit cell positions $\{\vec{\delta}_{\alpha}\}$ for which Eq.~\eqref{eq.Ds_qm} holds always exist in a system with TRS and (2) these positions are such that ${\rm Tr}(\mathcal{M})$ is minimized.  

In order for Eq.~\eqref{eq.Ds_qm} to hold, we need $D_s=(e^2/V\hbar^2)\partial^2\Omega/\partial q_i\partial q_j\big|_{\vec{q}=\vec{0}}$, which by Eq.~\eqref{eq.sfw_trs} requires that $({\rm d}_i\Delta^I)\partial^2_{\Delta^I}\Omega ({\rm d}_j\Delta^I)\big|_{\vec{q}=\vec{0}}=0$ for all $i,j$. As $\partial^2_{\Delta^I}\Omega$ is non-singular, this occurs if and only if ${\rm d}\Delta_{\alpha}/{\rm d}q_i\big|_{\vec{q}=\vec{0}}=0$ for all $\alpha$ and $i$. To show that positions where this is the case exist, we will use the following property of the order parameters: if the order parameters for a choice of intra-unit-cell positions $\{\vec{\delta}_{\alpha}\}$ are
$|\Delta_{\alpha}(\vec{q})| e^{i\theta_{\alpha}(\vec{q})}$, the order
parameters in the same model for another choice of positions
$\{\vec{\delta}_{\alpha}'\}$ are
$\Delta_{\alpha}' = |\Delta_{\alpha}|e^{i\theta_{\alpha}'}$, where
$\theta_{\alpha}' = \theta_{\alpha} - 2\vec{q}\cdot
(\vec{\delta}_{\alpha}' - \vec{\delta}_{\alpha})$
 (for a proof, see Appendix B in~\cite{Huhtinen2022}). 

Since the derivatives of the order parameters are purely imaginary at $\vec{q}=\vec{0}$,
\begin{equation}
    \frac{{\rm d}\Delta_{\alpha}}{{\rm d}q_i}\bigg|_{\vec{q}=\vec{0}} = i|\Delta_{\alpha}|\frac{{\rm d}\theta_{\alpha}}{{\rm d}q_i}\bigg|_{\vec{q}=\vec{0}}.
\end{equation}
The derivatives of $\Delta_{\alpha}$ and $\Delta_{\alpha}'$ as defined above are thus related by
\begin{equation}
  \frac{{\rm d}\Delta_{\alpha}^{I}}{{\rm d}q_j}\bigg|_{\vec{q}=\vec{0}}
  = \frac{{\rm d}(\Delta_{\alpha}')^{I}}{{\rm d}q_j}\bigg|_{\vec{q}=\vec{0}} +
  2i\Delta_{\alpha}[\vec{\delta}_{\alpha}'-\vec{\delta}_{\alpha}]_j, \label{eq.app_rel} 
\end{equation}
where $[\vec{\delta}_{\alpha}'-\vec{\delta}_{\alpha}]_j$ is the $j$:th component of the translation vector $\vec{\delta}_{\alpha}'-\vec{\delta}_{\alpha}$. 

If we know the derivatives for some arbitrary set of positions $\{\vec{\delta}_{\alpha}^0\}$, we can thus solve positions $\{\vec{\delta}_{\alpha}\}$ for which ${\rm d}_i\Delta^I\big|_{\vec{q}=\vec{0}}=0$ from Eq.~\eqref{eq.app_rel}. For any orbital where $\Delta_{\alpha}\neq 0$, they are given by
\begin{equation}
    [\vec{\delta}_{\alpha}]_i = [\vec{\delta}_{\alpha}^0]_i +
  \frac{1}{2}\frac{{\rm d}\theta_{\alpha}^{0}}{{\rm
      d}q_i}\bigg|_{\vec{q}=\vec{0}}, \label{eq.orb_shift}
\end{equation}
where $\theta_{\alpha}^0$ are the phases of the order parameters for orbital positions $\{\vec{\delta}_{\alpha}^0\}$. When $\Delta_{\alpha}=0$, the derivative is trivially zero, and the choice of position for that orbital is arbitrary. 

We have now found a choice of orbital positions for which $[D_s]_{ij} = (e^2/V\hbar^2) \partial^2\Omega/\partial q_i\partial q_j\big|_{\vec{q}=\vec{0}}$, and the quantum metric $\mathcal{M}$ is indeed related to the superfluid weight via Eq.~\eqref{eq.Ds_qm}. However, to determine $\{\vec{\delta}_{\alpha}\}$ from Eq.~\eqref{eq.orb_shift}, we still need knowledge of the derivatives of the order parameters for some initial choice of positions. In order to remove this requirement, we use the inequality
\begin{equation}
\frac{\partial^2\Omega}{\partial q_i^2}\bigg|_{\vec{q}=\vec{0}} \geq\frac{{\rm d}^2F}{{\rm d}q_i^2}\bigg|_{\vec{q}=\vec{0}},
\end{equation}
mentioned in Sec.~\ref{sec.mf_sfw}. This is a consequence of the positive definiteness of the Hessian matrix $\partial_{\Delta^I}^2\Omega$. The partial derivative $\partial^2\Omega/\partial q_i\partial q_j\big|_{\vec{q}=\vec{0}}$ is proportional to $\mathcal{M}_{ij}$, and the superfluid weight is thus directly related to the quantum metric whenever this inequality is saturated.  This is clearly a minimum of $\partial^2\Omega/\partial q_i^2\big|_{\vec{q}=\vec{0}}$, and thus of the diagonal components of the quantum metric. The orbital positions found previously are therefore such that they minimize the trace of the integrated quantum metric $\mathcal{M}$ over all possible choices of positions: the quantum metric related to the superfluid weight is the {\it minimal} quantum metric, a single-particle quantity which can be computed without solving the BCS ground state of the system. The minimization here is performed for the full integrated quantum metric, and not pointwise for each $\vec{k}$. 

It is important to remember here that the superfluid weight is not affected by the choice of intra-unit cell orbital positions in Eq.~\eqref{eq.fourier}, and the correct superfluid weight in a system with TRS can always be computed from Eq.~\eqref{eq.sfw_trs}, or equivalently Eq.~\eqref{eq.linrespresult}. The choice of orbital positions becomes crucial only when we want to relate the quantum metric to $D_s$, or use $D_s=(e^2/V\hbar^2)\partial^2\Omega/\partial q_i\partial q_j\big|_{\vec{q}=\vec{0}}$.

The above results are valid in systems with TRS. In the absence of TRS, the inequality Eq.~\eqref{eq.inequality} may not hold when ${\rm d}\mu/{\rm d}q_i\big|_{\vec{q}=\vec{0}}\neq 0$: then, the full Hessian matrix $\partial^2_{\Delta,\mu}\Omega$ defined in Eq.~\eqref{eq.hess} does not need to be positive semidefinite, because the $\mu$ for which the correct particle number $N$ is achieved is not a minimum. When ${\rm d}\mu/{\rm d}q_i\big|_{\vec{q}=\vec{0}}=0$, the terms involving the chemical potential can be ignored, and we recover a positive semidefinite Hessian matrix. Then, although Eq.~\eqref{eq.inequality} holds, the inequality may not be saturated for any choice of orbital positions, in contrast to a system with TRS. Indeed, ${\rm d}\Delta_{\alpha}/{\rm d}q_i\big|_{\vec{q}=\vec{0}}$ does not need to be purely imaginary in a system without TRS. The real part of these derivatives cannot be set to zero by only manipulating the phases of the order parameters with a shift of the orbital positions.
	
\bibliographystyle{varenna}
\bibliography{varenna_lecture_notes}
\end{document}